\documentclass[prl,aps,twocolumn,showpacs,floatfix,longbibliography]{revtex4-2}

\usepackage{mathrsfs}
\usepackage{rotating}
\usepackage{color}
\usepackage{graphicx}
\usepackage{dcolumn}
\usepackage{bm}
\usepackage{hyperref}
\usepackage[latin1]{inputenc}
\usepackage{pstricks}
\usepackage{amsmath,amssymb}
\usepackage{version}
\usepackage{epstopdf}
\DeclareGraphicsExtensions{.eps,.ps}

\usepackage{ulem}
\usepackage{float}
\usepackage{booktabs}
\usepackage{subfigure}
\usepackage{graphicx}
\usepackage{dcolumn}
\usepackage{bm}
\usepackage{ulem}
\usepackage{setspace}

\begin{document}

\title{Photoelectron Polarization Vortexes in Strong-Field Ionization}

\author{Pei-Lun He}\email{peilun@mpi-hd.mpg.de}
\author{Zhao-Han Zhang}
\email{zhao-han.zhang@mpi-hd.mpg.de}
\author{Karen Z. Hatsagortsyan}\email{k.hatsagortsyan@mpi-hd.mpg.de}
\author{Christoph H. Keitel} 
\affiliation{Max-Planck-Institut f\"ur Kernphysik, Saupfercheckweg 1, 69117 Heidelberg, Germany}

\date{\today}

\begin{abstract}
The spin polarization of photoelectrons induced by an intense linearly polarized laser field is investigated using numerical solutions of the time-dependent Schr\"odinger equation in companion with our analytic treatment via the spin-resolved strong-field approximation and classical trajectory Monte Carlo simulations.
We demonstrate that, even though the total polarization vanishes upon averaging over the photoelectron momentum, momentum-resolved spin polarization is significant, typically exhibiting a vortex structure relative to the laser polarization axis.
The polarization arises from the transfer of spin-orbital coupling in the bound state to the spin-correlated quantum orbits in the continuum.
The rescattering of photoelectrons at the atomic core plays an important role in forming the polarization vortex structure, while there is no significant effect of the spin-orbit coupling during the continuum dynamics.
Furthermore, spin-polarized electron holography is demonstrated, feasible for extracting fine structural information about the atom.
\end{abstract}

\maketitle

Spin-$1/2$ is an inherent property of an electron, providing a valuable signal for applications across various branches of physics.
For instance, in scattering experiments, the spin polarization can provide independent information on the molecular structure in addition to the electron angular distribution  \cite{kessler1969electron}. Spin and angle-resolved photoemission spectroscopy is a powerful tool for directly measuring the spin texture of electronic states \cite{lv2019angle}. Furthermore, polarized electrons can provide stringent tests of the standard model and diagnose new physics \cite{moortgat2008polarized}.

Ionization can serve as a source of spin-polarized electrons. Spin-orbit coupling in an atomic bound state can create a strong correlation between the electron's spin and orbital angular momentum. The correlation, combined with the significant angular momentum-dependent ionization probability, results in photoelectron polarization.
Thus, in single-photon and multiphoton ionization, nearly $100\%$ spin polarization can be achieved near the Cooper minimum \cite{fano1969spin, lambropoulos1973spin}. In tunneling ionization, circular dichroism due to nonadiabatic effects can lead to spin polarization of photoelectrons of noble gas atoms  \cite{barth2011nonadiabatic, barth2013nonadiabatic, barth2013spin,Barth_2014, liu2018deformation, eckart2018ultrafast,hu2023effect}. As a consequence, significant photoelectron polarization is feasible both when the correlated ion state is resolved and when sampling over all possible ion states \cite{hartung2016electron, liu2018energy, trabert2018spin}. In the multiphoton ionization regime, intermediate resonances play a crucial role in circular dichroism, resulting in photoelectrons with spin polarization that is sensitive to the photon energy \cite{Zhu_2016,Walker_2021}.
In the interaction between ultrastrong laser fields and highly charged ions (HCI), an electron spin-flip is possible during tunneling \cite{Klaiber_2014, Klaiber_2015}.

In strong-field ionization, photoelectrons can be driven back by the laser field to collide with their parent ions~\cite{corkum1993plasma}. The influence of spin-orbit coupling is significant in the bound state due to the degenerate atomic energy levels. The spin-singlet and spin-triplet states have distinct spectra in high-order above-threshold ionization \cite{Zille_2017}. 
While spin-orbit coupling at recollisions belongs to a high-order relativistic correction, it can yield a noticeable contribution in a weakly relativistic regime with HCI \cite{Walser_1999, Hu_1999, Walser_2002}. 
Recently, nonnegiligible contributions of spin-orbit coupling in the continuum are found in the case of large momentum transfer at moderate laser intensities with mid-infrared lasers \cite{maxwell2023relativistic}.
Beyond the single particle picture, spin-orbit coupling can drive the hole dynamics in the ion leading to time-dependent hole polarization \cite{rohringer2009multichannel,barth2014hole,kubel2019spatiotemporal,mayer2022role,stewart2023attosecond}. 
Recently, with the method of time-dependent configuration-interaction singles \cite{carlstrom2022general,carlstrom2022general2}, Carlstr\"om \textit{et al.} reported nontrivial spin polarization of photoelectrons ionized by a linearly polarized pulse when the correlated ion state is resolved \cite{carlstrom2023control}. The spin-orbit coupling in the electron's inelastic rescattering by the hole, combined with time-dependent spin dynamics of the latter, is responsible for this spin effect.
While the conservation law of angular momentum implies that net electron polarization is impossible when ionizing the spinless ground state of rare gas atoms with a linearly polarized laser field, angle-resolved spin polarization can still occur.

In this letter, we explore spin polarization in strong-field ionization induced by a linearly polarized laser pulse.
We have conducted a spin-resolved study of the ionization process both analytically, employing the strong field approximation (SFA), and numerically through the solution of the time-dependent Schr\"odinger equation (TDSE) and classical trajectory Monte Carlo (CTMC) simulations.
In contrast to a circularly polarized pulse, the total polarization of the ionized electron vanishes here. However, a strong entanglement of the angular distribution with the electron spin emerges, which arises from the spin-orbit coupling in the bound state that establishes a correlation between the orbital angular momentum and the spin of the valence shell electron. The correlation extends to the spin and the initial transverse velocity of the photoelectron at the tunnel exit, giving rise to the emergence of spin-dependent quantum orbits. Finally, the forward rescattering of the spin-dependent trajectories results in momentum-resolved spin polarization, exhibiting a vortex structure. While rescattering is an important ingredient in creating the considered spin effects, spin-orbit coupling at the rescattering plays a minor role.

Starting with our analytical approach, the spin-resolved wave function of the bound state, which accounts for the spin-orbit coupling, is given by
\begin{equation}
\psi_{\kappa j l m}(\mathbf{x})=2C_{\kappa l}\kappa^{3/2}(\kappa r)^{\nu-1}e^{-\kappa r }\Omega_{jlm}(\mathbf{x}), \label{wavefunction}
\end{equation}
where $\Omega_{jlm}(\mathbf{x})$ is the spinor spherical harmonics, $(j, m)$ the total angular quantum number, $\nu$ the effective principle quantum number, $I_p$ the ionization potential, and $\kappa=\sqrt{2I_p}$.
We adopt the strong field approximation \cite{keldysh1965ionization,faisal1973multiple,reiss1980effect}, which is valid in the regime of tunneling ionization, with which the two-component ionization amplitude reads \cite{supp}
\begin{equation}
\begin{aligned}
M(\mathbf{p};j,m)
=-i\int_{t_0}^{t_f} dt\begin{pmatrix}
 \langle \mathbf{p}_V (t) |\mathbf{x}\cdot\mathbf{E}(t)  |\chi_u(t)\rangle\\ 
 \langle \mathbf{p}_V (t) |\mathbf{x}\cdot\mathbf{E}(t)|\chi_d(t)\rangle
\end{pmatrix},
\end{aligned}\label{M1}
\end{equation}
where $|\mathbf{p}_V (t)\rangle $ is the Volkov wave function \cite{becker2002above}, $\mathbf{A}(t) = -A_0\textbf{e}_x\sin(\omega t+\phi)f(t)$ the vector potential, $f(t)$ the pulse envelope, $A_0$ the amplitude, $\omega$ the laser frequency, $\phi$ the carrier-envelope phase (CEP), $\mathbf{E}(t) = -\partial_t \mathbf{A}(t)$ the electric field, and $\chi_u$ and $\chi_d$ are the upper and lower components of the spinor, respectively.
The derivation details are included in the Supplementary Materials (SM) \cite{supp}.

In typical ionization experiments, the correlated ion information is not available. Therefore, summing over the magnetic quantum numbers is necessary
\begin{equation}
\left\langle \boldsymbol{\zeta}(\mathbf{p};j) \right\rangle = \frac{\sum_{m} M^\dagger (\mathbf{p};j,m)\boldsymbol{\sigma} M(\mathbf{p};j,m)}{\sum_{m}M^\dagger(\mathbf{p};j,m) M(\mathbf{p};j,m)},\label{polarization}
\end{equation}
where $\boldsymbol{\sigma}$ represents the Pauli matrices.
Since spin-orbit coupling belongs to high-order relativistic corrections, the upper and lower components of the spinor effectively decouple when starting with an initial state that has the eigenvalue $(j, m)$ of the total angular momentum operator \cite{Walser_1999, Hu_1999, Walser_2002}.
Thus, we can calculate the spin polarization from the numerical TDSE simulations, with the weight given by the Clebsch-Gordan coefficients \cite{sakurai2020modern}.
In this paper, we use atomic units and adopt the $z$-axis as the quantization axis for angular momentum, while the $x$-axis is the polarization axis.

\begin{figure}
\centering
\includegraphics[width=0.5\textwidth]{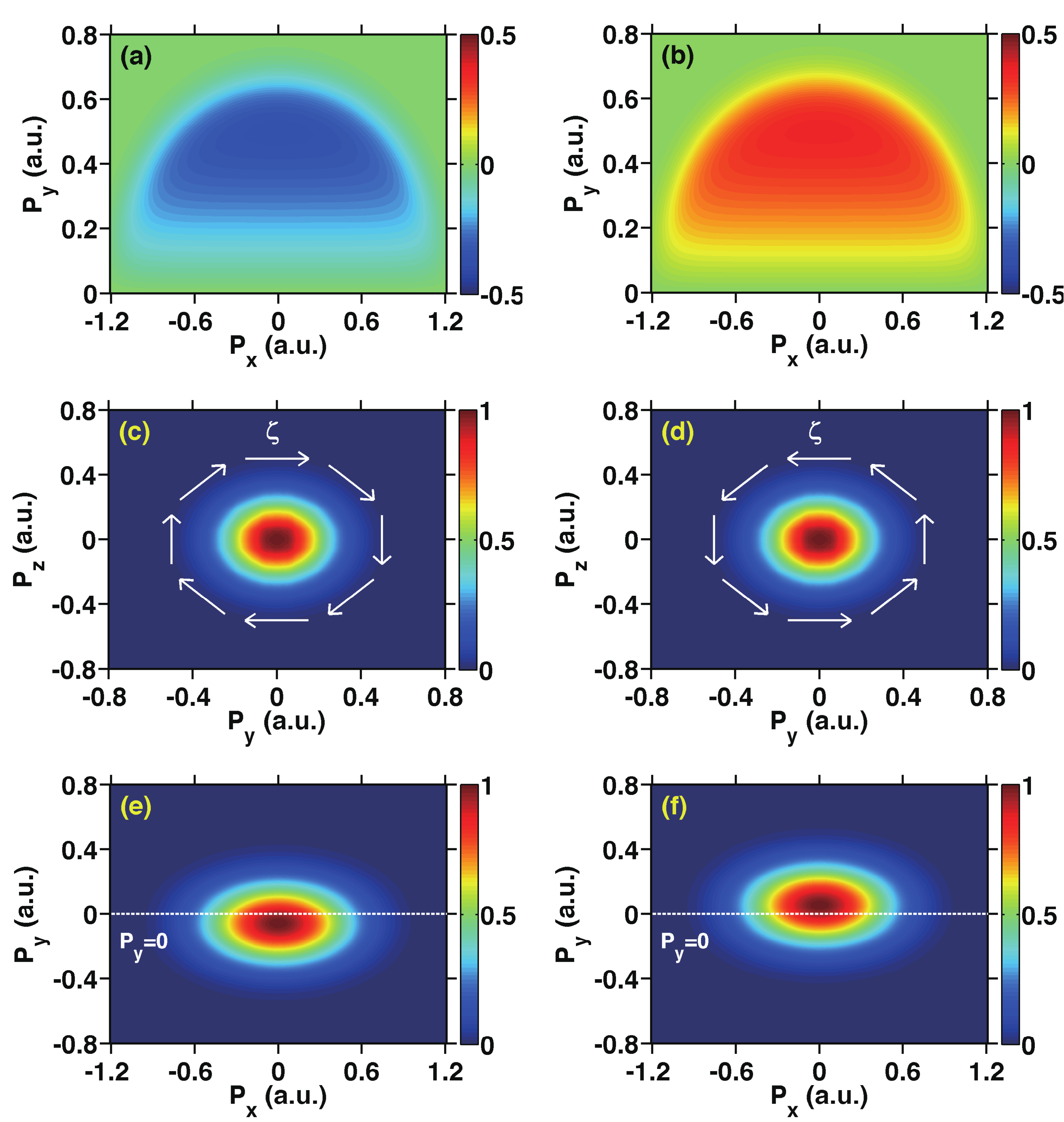}
  \caption{Time-resolved strong-field ionization of direct electrons from the $2\textup{P}_{\frac{3}{2}}$ state of He$^+$ via SFA. (Left column) Contribution from $E_x(t)>0$; (Right column) Contribution from $E_x(t)<0$. (a, b) Spin polarization $\zeta_z$ in the $x$-$y$ plane.
(c,~d) Photoelectron momentum distribution in the $y$-$z$ plane, with the white dashed line illustrating vortically polarized spin polarization.
(e, f) Corresponding photoelectron momentum distribution contributed from the $\chi^{(+)}$ orbit, where the white dashed line denotes the position of $p_y=0$. The laser wavelength is 800 nm with an intensity of $I= 10^{14}$ W/cm$^2$.
}
\label{figM1}
 \end{figure}

For a clear and unambiguous theoretical analysis, we first study ionization from the $2\textup{P}_{\frac{3}{2}}$ state of He$^+$ as a prototype. Utilizing the saddle point approximation for the time integration in Eq.~(\ref{M1}) alongside with Eq.~(\ref{polarization}), we derive the spin polarization at the tunneling exit as  
\begin{equation}
\left \langle \boldsymbol{\zeta}(\mathbf{p};j=\frac{3}{2}) \right \rangle \approx \frac{\mathbf{p} \times \mathbf{E}(t_r)}{\kappa |\mathbf{E}(t_r)|},\label{PolarizationM1}
\end{equation}
where $t_r$ represents the ionization time.
Equation (\ref{PolarizationM1}) illustrates the vortex structure of electron polarization relative to the laser polarization axis. The electron polarization at the tunneling exit is associated with a specific electron trajectory, and when combined with the ionization probability, it can be utilized in CTMC simulations to reveal momentum-resolved spin polarization. We neglect the spin precession in the continuum as it is a high-order relativistic correction \cite{bargmann1959precession}.

\begin{figure}
\centering
\includegraphics[width=0.5\textwidth]{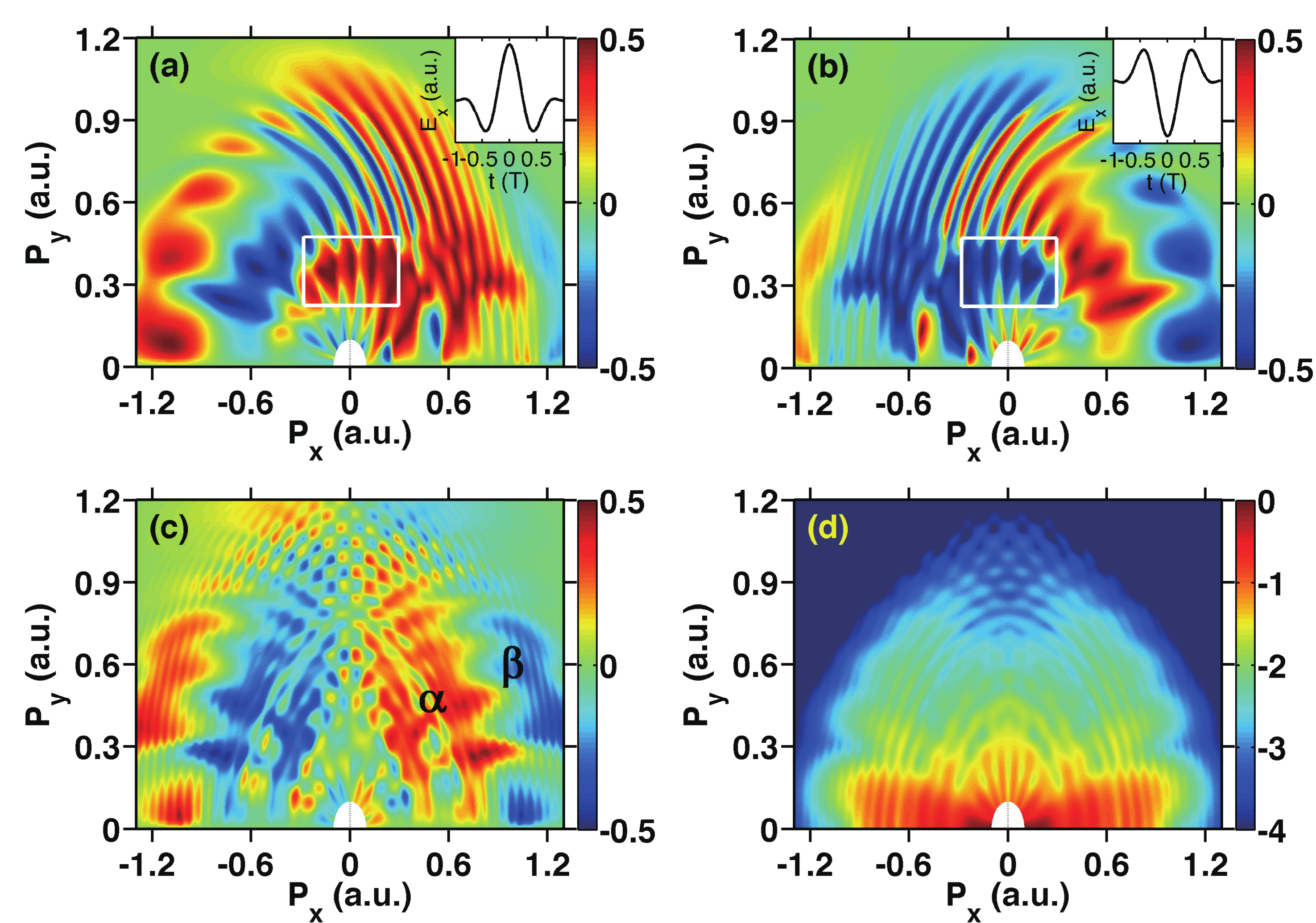}
  \caption{The spin polarization $\zeta_z$ in the $x-y$ plane obtained from TDSE simulations when the CEP of the  few-cycle pulse is (a) $\phi=0$, (b) $\phi=\pi$, and (c) $\phi$-averaged over $[0,2\pi]$.
The inset in (a, b) illustrates the corresponding electric field of the laser pulse, and the white boxes indicate the polarization contributed by the direct electron.
Panel (d) is the photoelectron momentum distribution corresponding to the CEP-averaged case.
The laser pulse has a wavelength of 800 nm and an intensity of $I= 10^{14}$ W/cm$^2$.
The pulse envelope is $f(t)=\cos^2(\frac{\pi t}{L})$, where the pulse duration is $L=2T$ for (a, b) and $L=4T$ for (c, d).
In panel (c), $\alpha$ and $\beta$ label regions where the polarization is positive and negative, respectively.
}
\label{figL}
\end{figure}

Due to the rotational symmetry with respect to the laser polarization axis,
it suffices to calculate the $z$ component of the polarization in the upper $x-y$ plane using Eq.~(\ref{wavefunction}) and Eq.~(\ref{polarization}), which reads 
\begin{equation}
\left \langle \zeta_z(\mathbf{p};j=\frac{3}{2}) \right \rangle
\approx  \frac{1}{2}\frac{ \left | \chi^{(+)}\right |^2-\left |\chi^{(-)}\right |^2}{\left | \chi^{(+)}\right |^2+\left |\chi^{(-)}\right |^2},\label{j32}
\end{equation}
where $\chi^{(\pm)}$ represents the wave function with orbital magnetic quantum numbers $m_l=\pm 1$.
The direct electron is negatively polarized in the upper $x-y$ plane when the electric field is positive [Fig.~\ref{figM1}(a)], and positively polarized when the electric field is negative [Fig.~\ref{figM1}(b)]. 
The $z$-component of the polarization shown in the upper $x-y$ plane implies a vortex structure in the $y-z$ plane [Figs.~\ref{figM1}(c,d)], whose origin has a simple explanation.
As Eq.~(\ref{j32}) indicates, the electron polarization $\zeta_z=\pm \frac{1}{2}$ is correlated with the $m_l=\pm 1$ orbital. The latter generates the most probable tunnel-ionized electron with a nonvanishing initial transverse velocity of $p^{(i)}_y = \mp E_x(t_r)/\kappa^2$ [Figs.~\ref{figM1}(e,f)],  according to the Perelomov-Popov-Terent'ev (PPT) ionization rate \cite{perelomov1966ionization, delone1991energy}:
\begin{equation}
\begin{aligned}
\Gamma^{(\pm)} \left(t_r,{p}_y,{p}_z\right) &\propto   \left(\frac{p_y}{\kappa}\mp\textup{sign}(E_x(t_r))\right)^2 
\exp\left[-\frac{\kappa p_y^2}{\left |E_x(t_r)  \right |}\right] .
\end{aligned}\label{rate}
\end{equation}
The initial transverse velocity is nonvanishing even in the adiabatic limit \cite{supp}, thus the existence of the photoelectron polarization vortex does not rely on nonadiabatic effects, in contrast to the case of circular polarization.

The contributions of the direct electron to the polarization cancel out in a monochromatic laser wave due to its symmetry in each half cycle. However, observing the spin polarization of the direct electron is possible with a few-cycle pulse, as illustrated by our TDSE simulation in Fig.~\ref{figL}. 
When the CEP is $\phi=0$, the polarization of the direct electrons (inside the white box) resembles that in Fig.~\ref{figM1}(a); conversely, at $\phi=\pi$, it resembles Fig.~\ref{figM1}(b).
Observing the polarization of the direct electron is more efficient using a few-cycle pulse with a longer wavelength and weaker intensity \cite{supp}.
Some regions in Figs.~\ref{figL}(a,b) are insensitive to variations in the CEP. Even after averaging over the CEP, significant polarization persists in Fig.~\ref{figL}(c), which also remains insensitive to the pulse duration.
The polarization in the first quadrant of Fig.~\ref{figL}(c) can be divided into two distinct regimes, labeled as $\alpha$ and $\beta$. 
Analyzing the corresponding photoelectron momentum distribution in Fig.~\ref{figL}(d), we can see that the $\alpha$ regime exhibits a dominant probability, while the $\beta$ regime has a much smaller probability and an energy larger than $2U_p$, where $U_p= A_0^2/4$ is the ponderomotive energy.

\begin{figure}
\centering
\includegraphics[width=0.5\textwidth]{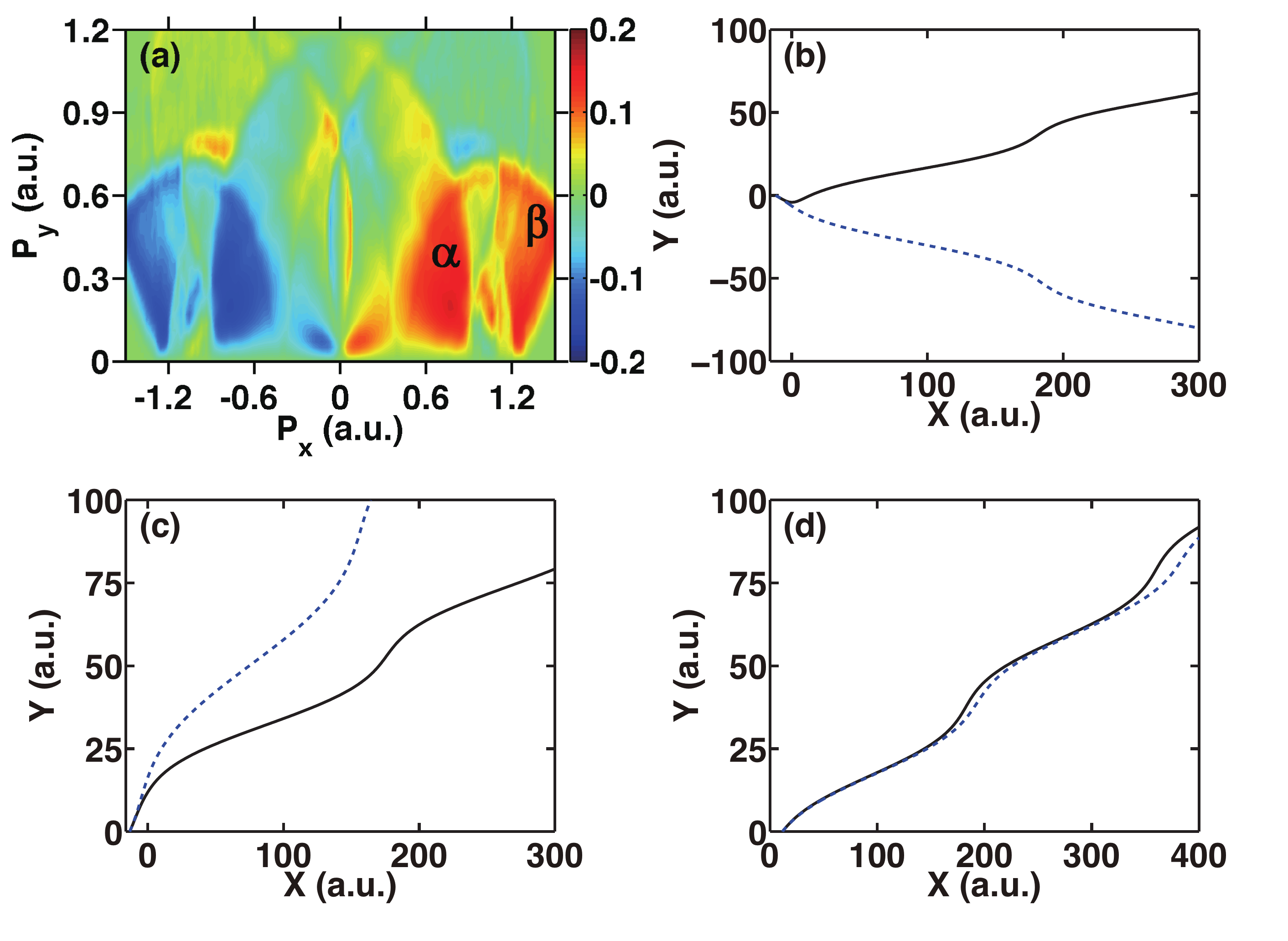}
  \caption{CTMC analysis of the spin polarization. 
(a) Spin polarization $\zeta_z$ in the $x-y$ plane.
(b-d) Trajectories that dominate the polarization in the first quadrant.
The blue dashed line is the trajectory without the continuum Coulomb force, while the black solid line is the trajectory including the continuum Coulomb force.
The laser pulse has a wavelength of 800 nm and an intensity of $I= 10^{14}$ W/cm$^2$.
}
\label{figCTMC}
\end{figure}

We have demonstrated that the momentum-resolved electron polarization in multicycle laser pulses is nonvanishing. It is primarily due to forward scattering at the recollision. To illustrate this, we carry out spin-resolved CTMC simulations in Fig.~\ref{figCTMC}(a), where the polarization in the $\alpha$ regime is correctly reproduced.
The typical trajectories that contribute to the first quadrant are shown in Figs.~\ref{figCTMC}(b-d).
Panels (c) and (d) are direct ionization trajectories, with negative and positive contributions to the polarization, respectively. When the continuum Coulomb force is neglected, their contributions cancel each other out. 
Panel (b) is the trajectory contributed by rescattering. It is ionized at instants when $E_x(t_r)>0$ and has a negative initial transverse velocity. Consequently, it has a positive contribution to the polarization.
Upon recollision, the transverse velocity changes sign and is scattered to the first quadrant, significantly enhancing photoelectron yields around $2U_p$ \cite{keil2016laser,he2018high}.
Therefore, the positive polarization observed in the $\alpha$ and $\beta$ regions in panel (a) originates from the classical mechanism.
However, in the $\beta$ regime, the quantum effect of the destructive interference of trajectories distort the result. We analyze the phase factors of the quantum orbits
 \cite{li2014classical,PhysRevA.92.043407,PhysRevA.100.053411}:
\begin{equation}
S=\int_{t_r}^{\infty}dt~\left[\frac{1}{2}\left(\mathbf{p}_c(t)+\mathbf{A}(t)\right)^2-\frac{2Z}{\left | \mathbf{r}_c(t) \right |}\right],
\end{equation}
where $\left(\mathbf{r}_c(t),\mathbf{p}_c(t)\right)$  is the classic trajectory.
In the $\beta$ regime, the phase shift due to the Coulomb potential leads to destructive interference between (b) and (d). Consequently, (c) dominates, resulting in a negative polarization \cite{supp}. 

To better understand the role of the Coulomb field in quantum interference, we solve the 2-dimensional TDSE in the integral form
\begin{equation}
|\Psi(t)\rangle=-i \int_{t_{\mathrm{i}}}^t d \tau U_\sigma(t, \tau) \mathbf{x} \cdot \mathbf{E}(\tau) |\Phi_0(\tau)\rangle,
\end{equation}
where additional insight can be gained by using a Yukawa potential with a variable force range in the evolution operator \cite{he2015photoelectron}
\begin{equation}
U_\sigma(t_1, t_2)=\mathcal{T} \exp \left[-i \int_{t_2}^{t_1}d t \left(\frac{1}{2}\textbf{p}^2 + \mathbf{x} \cdot \mathbf{E}(t) -\frac{Z}{r}e^{-\sigma r} \right) \right].
\end{equation}
The obtained polarization closely resembles the CTMC simulations in Fig.~\ref{figCTMC}(a) at the applied shortest force range with $\sigma=0.04$, when the force range is sufficiently long to encompass the laser-driven quiver radius of the photoelectron.
However, increasing the force range alters the polarization in the $\beta$ region, gradually causing the positive polarization to change to negative one similar to Fig.~\ref{figL}(c) \cite{supp}. 
The phase shift  due to the long-range tail of the Coulomb potential between orbitals with different $m_l$ is responsible for the sign change of the polarization, which is evident when examining the $y$-component of the polarization vortex
\begin{equation}
\left \langle \zeta_y(\mathbf{p};j=\frac{3}{2}) \right \rangle
\approx  -\frac{ \textup{Im} ~\chi^{(0)*}\chi^{(x)}}{\left | \chi^{(+)}\right |^2+\left |\chi^{(-)}\right |^2},
\end{equation}
where $\chi^{(x)}=\frac{1}{\sqrt{2}}(\chi^{(-)}-\chi^{(+)})$ and $\chi^{(0)}$ is the wave function with orbital magnetic quantum numbers $m_l=0$.
In this representation, it is the phase difference between the two orbits that leads to the observed polarization.
Altering the force range tunes the phase shift, thus leading to a change in the spin polarization.

\begin{figure}
\centering
\includegraphics[width=0.5\textwidth]{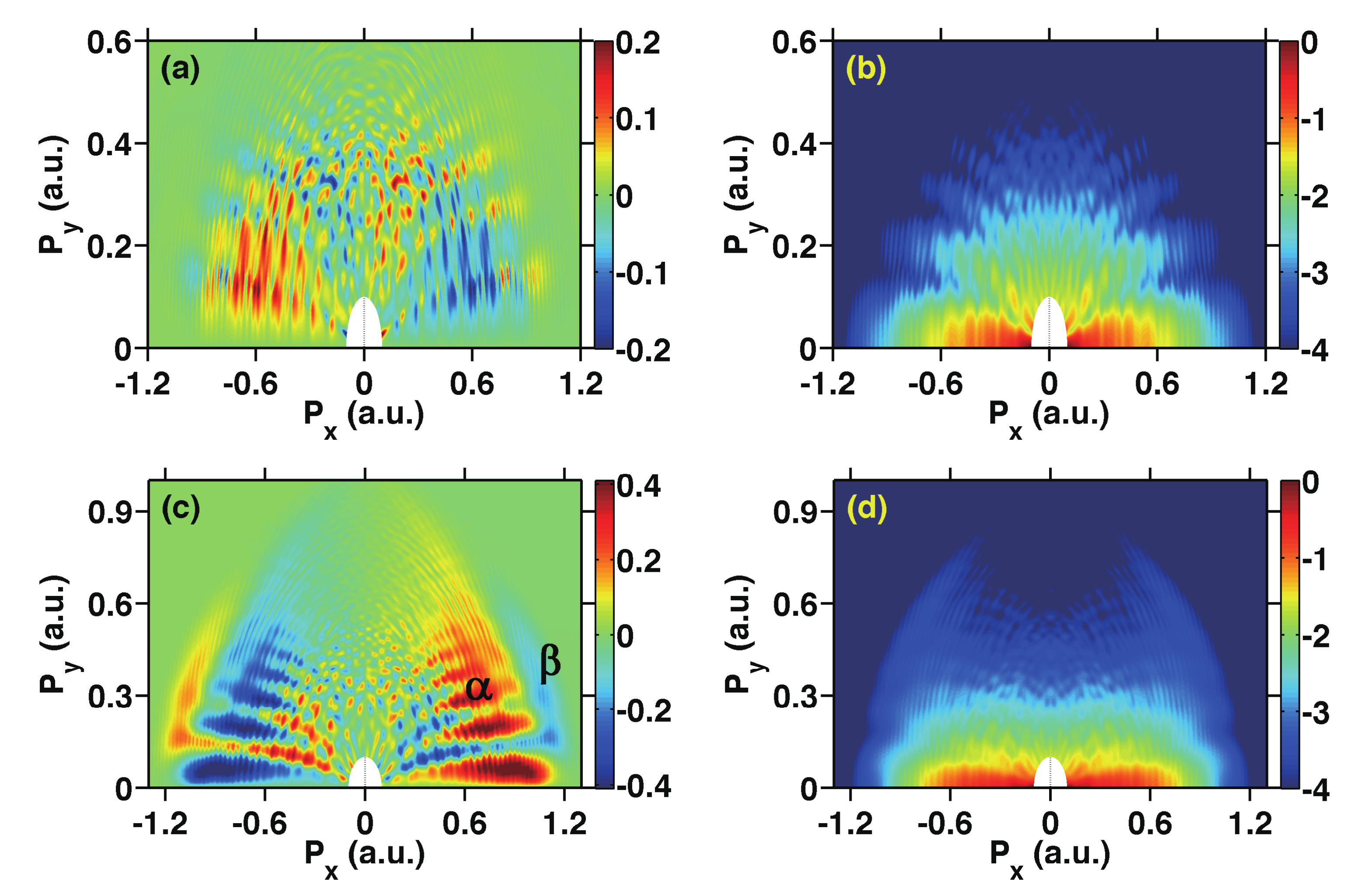}
  \caption{Spin-polarized photoelectron holography from TDSE simulations. 
Panels (a,c) display the spin-polarization $\zeta_z$. Panels (b,d) illustrate the photoelectron momentum distribution in the $x-y$ plane. Panels (a,b) correspond to the $5\textup{P}_{\frac{3}{2}}$ state of Xe, while panels (c,d) correspond to the $2\textup{P}_{\frac{3}{2}}$ state of He$^+$. The laser pulse has a wavelength of 2000 nm and the amplitude of the vector potential is $A_0=1$ a.u. 
The pulse envelope is $f(t)=\cos^2(\frac{\pi t}{L})$, where the pulse duration is $L=4T$.
}
\label{figHolography}
\end{figure}

Now, let us address the question of experimental observation of the considered polarization effect. In practice, various atomic shells could contribute to ionization, encompassing partial contributions from $ m_l=\pm 1$ states with different weights.
To observe the ionization signal predominantly from the valence shell, instead of applying the laser pulse with a wavelength of $800$ nm, one can choose a pulse with a smaller electric field, for instance, with $A_0=1$ a.u. and a longer wavelength of $2000$ nm. The characteristic photoelectron polarization region $\beta$, which reflects classical forward scattering and Coulomb long-range effects, is clearly visible for the $2\textup{P}_{\frac{3}{2}}$ state of He$^+$, as shown in Fig.~\ref{figHolography}(c).


Moreover, the interference between the direct and rescattered trajectories is known to lead to the  quantum effect of photoelectron holography \cite{huismans2011time,Huismans_2012,bian2012attosecond,Zhou_2016}.
The typical spider structure is visible in region $\alpha$ in the momentum distribution [Figs.~\ref{figHolography}(b,d)], with enhanced resolution in the spin polarization [Figs.~\ref{figHolography}(a,c)].
Hence, in addition to the momentum distribution, the holography spider structure also appears in the spin polarization, which we refer to as spin holography.
Not only is spin holography more sensitive to the atomic structure than the common strong field holography [cf. Fig.~\ref{figHolography}(d) with Fig.~\ref{figHolography}(c)], but also it can provide additional information. 
Spin holography can resolve the short-range component of the atomic potential, which alters both the classical dynamics and quantum phase shift, leading to the distinct difference in Figs.~\ref{figHolography}(a) and (c) for Xe and He$^+$.
Furthermore, spin holography is sensitive to the bound-state wave function, namely, to the spin-orbit coupling in the latter.
Apart from the classical action, the photoelectron picks up an additional phase at the tunneling exit \cite{liu2016phase} according to the orbital angular momentum of the initial state entangled with the spin, altering spin holography.
Furthermore, the polarizations contributed by the $\textup{P}_{\frac{3}{2}}$ and the $\textup{P}_{\frac{1}{2}}$ shells have opposite signs. 
Thus, the polarization could resolve the weights of these two channels, providing information about the hole dynamics of the ion. 
To fully exploit the capabilities of spin holography, comprehensive analysis is necessary in the future.

In conclusion, our study demonstrates momentum-resolved spin polarization of photoelectrons during strong-field ionization in a linearly polarized laser field, where a vanishing average polarization is commonly known. The vortex structure of the polarization relative to the laser's polarization axis is attributed to the forward rescattering of the polarization-correlated quantum orbits and the influence of the long tail of the atomic potential.
The polarization is pronounced and is experimentally accessible using mid-infrared laser fields interacting with rare gas atoms. 
We highlight the notion of photoelectron spin holography which can provide extra structural information about the atom.
Investigations on the role of nondipole effects and spin-orbit coupling are underway.


P.-L. H. thanks Fang Liu for an inspiring discussion on the effects of spin-orbit coupling in hole dynamics.

\bibliography{references.bib}

\begin{thebibliography}{53}%
\makeatletter
\providecommand \@ifxundefined [1]{%
 \@ifx{#1\undefined}
}%
\providecommand \@ifnum [1]{%
 \ifnum #1\expandafter \@firstoftwo
 \else \expandafter \@secondoftwo
 \fi
}%
\providecommand \@ifx [1]{%
 \ifx #1\expandafter \@firstoftwo
 \else \expandafter \@secondoftwo
 \fi
}%
\providecommand \natexlab [1]{#1}%
\providecommand \enquote  [1]{``#1''}%
\providecommand \bibnamefont  [1]{#1}%
\providecommand \bibfnamefont [1]{#1}%
\providecommand \citenamefont [1]{#1}%
\providecommand \href@noop [0]{\@secondoftwo}%
\providecommand \href [0]{\begingroup \@sanitize@url \@href}%
\providecommand \@href[1]{\@@startlink{#1}\@@href}%
\providecommand \@@href[1]{\endgroup#1\@@endlink}%
\providecommand \@sanitize@url [0]{\catcode `\\12\catcode `\$12\catcode
  `\&12\catcode `\#12\catcode `\^12\catcode `\_12\catcode `\%12\relax}%
\providecommand \@@startlink[1]{}%
\providecommand \@@endlink[0]{}%
\providecommand \url  [0]{\begingroup\@sanitize@url \@url }%
\providecommand \@url [1]{\endgroup\@href {#1}{\urlprefix }}%
\providecommand \urlprefix  [0]{URL }%
\providecommand \Eprint [0]{\href }%
\providecommand \doibase [0]{https://doi.org/}%
\providecommand \selectlanguage [0]{\@gobble}%
\providecommand \bibinfo  [0]{\@secondoftwo}%
\providecommand \bibfield  [0]{\@secondoftwo}%
\providecommand \translation [1]{[#1]}%
\providecommand \BibitemOpen [0]{}%
\providecommand \bibitemStop [0]{}%
\providecommand \bibitemNoStop [0]{.\EOS\space}%
\providecommand \EOS [0]{\spacefactor3000\relax}%
\providecommand \BibitemShut  [1]{\csname bibitem#1\endcsname}%
\let\auto@bib@innerbib\@empty
\bibitem [{\citenamefont {Kessler}(1969)}]{kessler1969electron}%
  \BibitemOpen
  \bibfield  {author} {\bibinfo {author} {\bibfnamefont {J.}~\bibnamefont
  {Kessler}},\ }\bibfield  {title} {\bibinfo {title} {Electron spin
  polarization by low-energy scattering from unpolarized targets},\ }\href@noop
  {} {\bibfield  {journal} {\bibinfo  {journal} {Reviews of Modern Physics}\
  }\textbf {\bibinfo {volume} {41}},\ \bibinfo {pages} {3} (\bibinfo {year}
  {1969})}\BibitemShut {NoStop}%
\bibitem [{\citenamefont {Lv}\ \emph {et~al.}(2019)\citenamefont {Lv},
  \citenamefont {Qian},\ and\ \citenamefont {Ding}}]{lv2019angle}%
  \BibitemOpen
  \bibfield  {author} {\bibinfo {author} {\bibfnamefont {B.}~\bibnamefont
  {Lv}}, \bibinfo {author} {\bibfnamefont {T.}~\bibnamefont {Qian}},\ and\
  \bibinfo {author} {\bibfnamefont {H.}~\bibnamefont {Ding}},\ }\bibfield
  {title} {\bibinfo {title} {Angle-resolved photoemission spectroscopy and its
  application to topological materials},\ }\href@noop {} {\bibfield  {journal}
  {\bibinfo  {journal} {Nature Reviews Physics}\ }\textbf {\bibinfo {volume}
  {1}},\ \bibinfo {pages} {609} (\bibinfo {year} {2019})}\BibitemShut {NoStop}%
\bibitem [{\citenamefont {Moortgat-Pick}\ \emph {et~al.}(2008)\citenamefont
  {Moortgat-Pick}, \citenamefont {Abe}, \citenamefont {Alexander},
  \citenamefont {Ananthanarayan}, \citenamefont {Babich}, \citenamefont
  {Bharadwaj}, \citenamefont {Barber}, \citenamefont {Bartl}, \citenamefont
  {Brachmann}, \citenamefont {Chen} \emph {et~al.}}]{moortgat2008polarized}%
  \BibitemOpen
  \bibfield  {author} {\bibinfo {author} {\bibfnamefont {G.}~\bibnamefont
  {Moortgat-Pick}}, \bibinfo {author} {\bibfnamefont {T.}~\bibnamefont {Abe}},
  \bibinfo {author} {\bibfnamefont {G.}~\bibnamefont {Alexander}}, \bibinfo
  {author} {\bibfnamefont {B.}~\bibnamefont {Ananthanarayan}}, \bibinfo
  {author} {\bibfnamefont {A.}~\bibnamefont {Babich}}, \bibinfo {author}
  {\bibfnamefont {V.}~\bibnamefont {Bharadwaj}}, \bibinfo {author}
  {\bibfnamefont {D.}~\bibnamefont {Barber}}, \bibinfo {author} {\bibfnamefont
  {A.}~\bibnamefont {Bartl}}, \bibinfo {author} {\bibfnamefont
  {A.}~\bibnamefont {Brachmann}}, \bibinfo {author} {\bibfnamefont
  {S.}~\bibnamefont {Chen}}, \emph {et~al.},\ }\bibfield  {title} {\bibinfo
  {title} {Polarized positrons and electrons at the linear collider},\
  }\href@noop {} {\bibfield  {journal} {\bibinfo  {journal} {Physics Reports}\
  }\textbf {\bibinfo {volume} {460}},\ \bibinfo {pages} {131} (\bibinfo {year}
  {2008})}\BibitemShut {NoStop}%
\bibitem [{\citenamefont {Fano}(1969)}]{fano1969spin}%
  \BibitemOpen
  \bibfield  {author} {\bibinfo {author} {\bibfnamefont {U.}~\bibnamefont
  {Fano}},\ }\bibfield  {title} {\bibinfo {title} {Spin orientation of
  photoelectrons ejected by circularly polarized light},\ }\href@noop {}
  {\bibfield  {journal} {\bibinfo  {journal} {Physical Review}\ }\textbf
  {\bibinfo {volume} {178}},\ \bibinfo {pages} {131} (\bibinfo {year}
  {1969})}\BibitemShut {NoStop}%
\bibitem [{\citenamefont {Lambropoulos}(1973)}]{lambropoulos1973spin}%
  \BibitemOpen
  \bibfield  {author} {\bibinfo {author} {\bibfnamefont {P.}~\bibnamefont
  {Lambropoulos}},\ }\bibfield  {title} {\bibinfo {title} {Spin-orbit coupling
  and photoelectron polarization in multiphoton ionization of atoms},\
  }\href@noop {} {\bibfield  {journal} {\bibinfo  {journal} {Physical Review
  Letters}\ }\textbf {\bibinfo {volume} {30}},\ \bibinfo {pages} {413}
  (\bibinfo {year} {1973})}\BibitemShut {NoStop}%
\bibitem [{\citenamefont {Barth}\ and\ \citenamefont
  {Smirnova}(2011)}]{barth2011nonadiabatic}%
  \BibitemOpen
  \bibfield  {author} {\bibinfo {author} {\bibfnamefont {I.}~\bibnamefont
  {Barth}}\ and\ \bibinfo {author} {\bibfnamefont {O.}~\bibnamefont
  {Smirnova}},\ }\bibfield  {title} {\bibinfo {title} {Nonadiabatic tunneling
  in circularly polarized laser fields: Physical picture and calculations},\
  }\href@noop {} {\bibfield  {journal} {\bibinfo  {journal} {Physical Review
  A}\ }\textbf {\bibinfo {volume} {84}},\ \bibinfo {pages} {063415} (\bibinfo
  {year} {2011})}\BibitemShut {NoStop}%
\bibitem [{\citenamefont {Barth}\ and\ \citenamefont
  {Smirnova}(2013{\natexlab{a}})}]{barth2013nonadiabatic}%
  \BibitemOpen
  \bibfield  {author} {\bibinfo {author} {\bibfnamefont {I.}~\bibnamefont
  {Barth}}\ and\ \bibinfo {author} {\bibfnamefont {O.}~\bibnamefont
  {Smirnova}},\ }\bibfield  {title} {\bibinfo {title} {Nonadiabatic tunneling
  in circularly polarized laser fields. ii. derivation of formulas},\
  }\href@noop {} {\bibfield  {journal} {\bibinfo  {journal} {Physical Review
  A}\ }\textbf {\bibinfo {volume} {87}},\ \bibinfo {pages} {013433} (\bibinfo
  {year} {2013}{\natexlab{a}})}\BibitemShut {NoStop}%
\bibitem [{\citenamefont {Barth}\ and\ \citenamefont
  {Smirnova}(2013{\natexlab{b}})}]{barth2013spin}%
  \BibitemOpen
  \bibfield  {author} {\bibinfo {author} {\bibfnamefont {I.}~\bibnamefont
  {Barth}}\ and\ \bibinfo {author} {\bibfnamefont {O.}~\bibnamefont
  {Smirnova}},\ }\bibfield  {title} {\bibinfo {title} {Spin-polarized electrons
  produced by strong-field ionization},\ }\href@noop {} {\bibfield  {journal}
  {\bibinfo  {journal} {Physical Review A}\ }\textbf {\bibinfo {volume} {88}},\
  \bibinfo {pages} {013401} (\bibinfo {year} {2013}{\natexlab{b}})}\BibitemShut
  {NoStop}%
\bibitem [{\citenamefont {Barth}\ and\ \citenamefont
  {Smirnova}(2014{\natexlab{a}})}]{Barth_2014}%
  \BibitemOpen
  \bibfield  {author} {\bibinfo {author} {\bibfnamefont {I.}~\bibnamefont
  {Barth}}\ and\ \bibinfo {author} {\bibfnamefont {O.}~\bibnamefont
  {Smirnova}},\ }\bibfield  {title} {\bibinfo {title} {Hole dynamics and spin
  currents after ionization in strong circularly polarized laser fields},\
  }\href@noop {} {\bibfield  {journal} {\bibinfo  {journal} {Journal of Physics
  B: Atomic, Molecular and Optical Physics}\ }\textbf {\bibinfo {volume}
  {47}},\ \bibinfo {pages} {204020} (\bibinfo {year}
  {2014}{\natexlab{a}})}\BibitemShut {NoStop}%
\bibitem [{\citenamefont {Liu}\ \emph {et~al.}(2018{\natexlab{a}})\citenamefont
  {Liu}, \citenamefont {Ni}, \citenamefont {Renziehausen}, \citenamefont
  {Rost},\ and\ \citenamefont {Barth}}]{liu2018deformation}%
  \BibitemOpen
  \bibfield  {author} {\bibinfo {author} {\bibfnamefont {K.}~\bibnamefont
  {Liu}}, \bibinfo {author} {\bibfnamefont {H.}~\bibnamefont {Ni}}, \bibinfo
  {author} {\bibfnamefont {K.}~\bibnamefont {Renziehausen}}, \bibinfo {author}
  {\bibfnamefont {J.-M.}\ \bibnamefont {Rost}},\ and\ \bibinfo {author}
  {\bibfnamefont {I.}~\bibnamefont {Barth}},\ }\bibfield  {title} {\bibinfo
  {title} {Deformation of atomic ${p}_{\ifmmode\pm\else\textpm\fi{}}$ orbitals
  in strong elliptically polarized laser fields: Ionization time drifts and
  spatial photoelectron separation},\ }\href
  {https://doi.org/10.1103/PhysRevLett.121.203201} {\bibfield  {journal}
  {\bibinfo  {journal} {Phys. Rev. Lett.}\ }\textbf {\bibinfo {volume} {121}},\
  \bibinfo {pages} {203201} (\bibinfo {year} {2018}{\natexlab{a}})}\BibitemShut
  {NoStop}%
\bibitem [{\citenamefont {Eckart}\ \emph {et~al.}(2018)\citenamefont {Eckart},
  \citenamefont {Kunitski}, \citenamefont {Richter}, \citenamefont {Hartung},
  \citenamefont {Rist}, \citenamefont {Trinter}, \citenamefont {Fehre},
  \citenamefont {Schlott}, \citenamefont {Henrichs}, \citenamefont {Schmidt}
  \emph {et~al.}}]{eckart2018ultrafast}%
  \BibitemOpen
  \bibfield  {author} {\bibinfo {author} {\bibfnamefont {S.}~\bibnamefont
  {Eckart}}, \bibinfo {author} {\bibfnamefont {M.}~\bibnamefont {Kunitski}},
  \bibinfo {author} {\bibfnamefont {M.}~\bibnamefont {Richter}}, \bibinfo
  {author} {\bibfnamefont {A.}~\bibnamefont {Hartung}}, \bibinfo {author}
  {\bibfnamefont {J.}~\bibnamefont {Rist}}, \bibinfo {author} {\bibfnamefont
  {F.}~\bibnamefont {Trinter}}, \bibinfo {author} {\bibfnamefont
  {K.}~\bibnamefont {Fehre}}, \bibinfo {author} {\bibfnamefont
  {N.}~\bibnamefont {Schlott}}, \bibinfo {author} {\bibfnamefont
  {K.}~\bibnamefont {Henrichs}}, \bibinfo {author} {\bibfnamefont {L.~P.~H.}\
  \bibnamefont {Schmidt}}, \emph {et~al.},\ }\bibfield  {title} {\bibinfo
  {title} {Ultrafast preparation and detection of ring currents in single
  atoms},\ }\href@noop {} {\bibfield  {journal} {\bibinfo  {journal} {Nature
  Physics}\ }\textbf {\bibinfo {volume} {14}},\ \bibinfo {pages} {701}
  (\bibinfo {year} {2018})}\BibitemShut {NoStop}%
\bibitem [{\citenamefont {Hu}\ \emph {et~al.}(2023)\citenamefont {Hu},
  \citenamefont {Liu}, \citenamefont {Renziehausen}, \citenamefont {Tian},
  \citenamefont {Zhang}, \citenamefont {Li}, \citenamefont {Zhou},\ and\
  \citenamefont {Lu}}]{hu2023effect}%
  \BibitemOpen
  \bibfield  {author} {\bibinfo {author} {\bibfnamefont {Y.}~\bibnamefont
  {Hu}}, \bibinfo {author} {\bibfnamefont {K.}~\bibnamefont {Liu}}, \bibinfo
  {author} {\bibfnamefont {K.}~\bibnamefont {Renziehausen}}, \bibinfo {author}
  {\bibfnamefont {Y.}~\bibnamefont {Tian}}, \bibinfo {author} {\bibfnamefont
  {Q.}~\bibnamefont {Zhang}}, \bibinfo {author} {\bibfnamefont
  {M.}~\bibnamefont {Li}}, \bibinfo {author} {\bibfnamefont {Y.}~\bibnamefont
  {Zhou}},\ and\ \bibinfo {author} {\bibfnamefont {P.}~\bibnamefont {Lu}},\
  }\bibfield  {title} {\bibinfo {title} {Effect of spin-orbit coupling in
  laser-induced ionization of atoms},\ }\href@noop {} {\bibfield  {journal}
  {\bibinfo  {journal} {Physical Review A}\ }\textbf {\bibinfo {volume}
  {108}},\ \bibinfo {pages} {023113} (\bibinfo {year} {2023})}\BibitemShut
  {NoStop}%
\bibitem [{\citenamefont {Hartung}\ \emph {et~al.}(2016)\citenamefont
  {Hartung}, \citenamefont {Morales}, \citenamefont {Kunitski}, \citenamefont
  {Henrichs}, \citenamefont {Laucke}, \citenamefont {Richter}, \citenamefont
  {Jahnke}, \citenamefont {Kalinin}, \citenamefont {Sch{\"o}ffler},
  \citenamefont {Schmidt} \emph {et~al.}}]{hartung2016electron}%
  \BibitemOpen
  \bibfield  {author} {\bibinfo {author} {\bibfnamefont {A.}~\bibnamefont
  {Hartung}}, \bibinfo {author} {\bibfnamefont {F.}~\bibnamefont {Morales}},
  \bibinfo {author} {\bibfnamefont {M.}~\bibnamefont {Kunitski}}, \bibinfo
  {author} {\bibfnamefont {K.}~\bibnamefont {Henrichs}}, \bibinfo {author}
  {\bibfnamefont {A.}~\bibnamefont {Laucke}}, \bibinfo {author} {\bibfnamefont
  {M.}~\bibnamefont {Richter}}, \bibinfo {author} {\bibfnamefont
  {T.}~\bibnamefont {Jahnke}}, \bibinfo {author} {\bibfnamefont
  {A.}~\bibnamefont {Kalinin}}, \bibinfo {author} {\bibfnamefont
  {M.}~\bibnamefont {Sch{\"o}ffler}}, \bibinfo {author} {\bibfnamefont
  {L.~P.~H.}\ \bibnamefont {Schmidt}}, \emph {et~al.},\ }\bibfield  {title}
  {\bibinfo {title} {Electron spin polarization in strong-field ionization of
  xenon atoms},\ }\href@noop {} {\bibfield  {journal} {\bibinfo  {journal}
  {Nature Photonics}\ }\textbf {\bibinfo {volume} {10}},\ \bibinfo {pages}
  {526} (\bibinfo {year} {2016})}\BibitemShut {NoStop}%
\bibitem [{\citenamefont {Liu}\ \emph {et~al.}(2018{\natexlab{b}})\citenamefont
  {Liu}, \citenamefont {Shao}, \citenamefont {Han}, \citenamefont {Ge},
  \citenamefont {Deng}, \citenamefont {Wu}, \citenamefont {Gong},\ and\
  \citenamefont {Liu}}]{liu2018energy}%
  \BibitemOpen
  \bibfield  {author} {\bibinfo {author} {\bibfnamefont {M.-M.}\ \bibnamefont
  {Liu}}, \bibinfo {author} {\bibfnamefont {Y.}~\bibnamefont {Shao}}, \bibinfo
  {author} {\bibfnamefont {M.}~\bibnamefont {Han}}, \bibinfo {author}
  {\bibfnamefont {P.}~\bibnamefont {Ge}}, \bibinfo {author} {\bibfnamefont
  {Y.}~\bibnamefont {Deng}}, \bibinfo {author} {\bibfnamefont {C.}~\bibnamefont
  {Wu}}, \bibinfo {author} {\bibfnamefont {Q.}~\bibnamefont {Gong}},\ and\
  \bibinfo {author} {\bibfnamefont {Y.}~\bibnamefont {Liu}},\ }\bibfield
  {title} {\bibinfo {title} {Energy-and momentum-resolved photoelectron spin
  polarization in multiphoton ionization of xe by circularly polarized
  fields},\ }\href@noop {} {\bibfield  {journal} {\bibinfo  {journal} {Physical
  Review Letters}\ }\textbf {\bibinfo {volume} {120}},\ \bibinfo {pages}
  {043201} (\bibinfo {year} {2018}{\natexlab{b}})}\BibitemShut {NoStop}%
\bibitem [{\citenamefont {Trabert}\ \emph {et~al.}(2018)\citenamefont
  {Trabert}, \citenamefont {Hartung}, \citenamefont {Eckart}, \citenamefont
  {Trinter}, \citenamefont {Kalinin}, \citenamefont {Sch{\"o}ffler},
  \citenamefont {Schmidt}, \citenamefont {Jahnke}, \citenamefont {Kunitski},\
  and\ \citenamefont {D{\"o}rner}}]{trabert2018spin}%
  \BibitemOpen
  \bibfield  {author} {\bibinfo {author} {\bibfnamefont {D.}~\bibnamefont
  {Trabert}}, \bibinfo {author} {\bibfnamefont {A.}~\bibnamefont {Hartung}},
  \bibinfo {author} {\bibfnamefont {S.}~\bibnamefont {Eckart}}, \bibinfo
  {author} {\bibfnamefont {F.}~\bibnamefont {Trinter}}, \bibinfo {author}
  {\bibfnamefont {A.}~\bibnamefont {Kalinin}}, \bibinfo {author} {\bibfnamefont
  {M.}~\bibnamefont {Sch{\"o}ffler}}, \bibinfo {author} {\bibfnamefont
  {L.~P.~H.}\ \bibnamefont {Schmidt}}, \bibinfo {author} {\bibfnamefont
  {T.}~\bibnamefont {Jahnke}}, \bibinfo {author} {\bibfnamefont
  {M.}~\bibnamefont {Kunitski}},\ and\ \bibinfo {author} {\bibfnamefont
  {R.}~\bibnamefont {D{\"o}rner}},\ }\bibfield  {title} {\bibinfo {title} {Spin
  and angular momentum in strong-field ionization},\ }\href@noop {} {\bibfield
  {journal} {\bibinfo  {journal} {Physical Review Letters}\ }\textbf {\bibinfo
  {volume} {120}},\ \bibinfo {pages} {043202} (\bibinfo {year}
  {2018})}\BibitemShut {NoStop}%
\bibitem [{\citenamefont {Zhu}\ \emph {et~al.}(2016)\citenamefont {Zhu},
  \citenamefont {Lan}, \citenamefont {Liu}, \citenamefont {Li}, \citenamefont
  {Liu}, \citenamefont {Zhang}, \citenamefont {Barth}, \citenamefont {Lu},
  \citenamefont {Moshammer}, \citenamefont {Feuerstein}, \citenamefont
  {Schmitt}, \citenamefont {Dorn}, \citenamefont {Schroter}, \citenamefont
  {Ullrich}, \citenamefont {Rottke}, \citenamefont {Trump}, \citenamefont
  {Wittmann}, \citenamefont {Korn}, \citenamefont {Hoffmann},\ and\
  \citenamefont {Sandner}}]{Zhu_2016}%
  \BibitemOpen
  \bibfield  {author} {\bibinfo {author} {\bibfnamefont {X.}~\bibnamefont
  {Zhu}}, \bibinfo {author} {\bibfnamefont {P.}~\bibnamefont {Lan}}, \bibinfo
  {author} {\bibfnamefont {K.}~\bibnamefont {Liu}}, \bibinfo {author}
  {\bibfnamefont {Y.}~\bibnamefont {Li}}, \bibinfo {author} {\bibfnamefont
  {X.}~\bibnamefont {Liu}}, \bibinfo {author} {\bibfnamefont {Q.}~\bibnamefont
  {Zhang}}, \bibinfo {author} {\bibfnamefont {I.}~\bibnamefont {Barth}},
  \bibinfo {author} {\bibfnamefont {P.}~\bibnamefont {Lu}}, \bibinfo {author}
  {\bibfnamefont {R.}~\bibnamefont {Moshammer}}, \bibinfo {author}
  {\bibfnamefont {B.}~\bibnamefont {Feuerstein}}, \bibinfo {author}
  {\bibfnamefont {W.}~\bibnamefont {Schmitt}}, \bibinfo {author} {\bibfnamefont
  {A.}~\bibnamefont {Dorn}}, \bibinfo {author} {\bibfnamefont {C.~D.}\
  \bibnamefont {Schroter}}, \bibinfo {author} {\bibfnamefont {J.}~\bibnamefont
  {Ullrich}}, \bibinfo {author} {\bibfnamefont {H.}~\bibnamefont {Rottke}},
  \bibinfo {author} {\bibfnamefont {C.}~\bibnamefont {Trump}}, \bibinfo
  {author} {\bibfnamefont {M.}~\bibnamefont {Wittmann}}, \bibinfo {author}
  {\bibfnamefont {G.}~\bibnamefont {Korn}}, \bibinfo {author} {\bibfnamefont
  {K.}~\bibnamefont {Hoffmann}},\ and\ \bibinfo {author} {\bibfnamefont
  {W.}~\bibnamefont {Sandner}},\ }\bibfield  {title} {\bibinfo {title}
  {{Helicity sensitive enhancement of strong-field ionization in circularly
  polarized laser fields}},\ }\href@noop {} {\bibfield  {journal} {\bibinfo
  {journal} {Optics Express}\ }\textbf {\bibinfo {volume} {24}},\ \bibinfo
  {pages} {4196} (\bibinfo {year} {2016})}\BibitemShut {NoStop}%
\bibitem [{\citenamefont {Walker}\ \emph {et~al.}(2021)\citenamefont {Walker},
  \citenamefont {Kolanz}, \citenamefont {Venzke},\ and\ \citenamefont
  {Becker}}]{Walker_2021}%
  \BibitemOpen
  \bibfield  {author} {\bibinfo {author} {\bibfnamefont {S.}~\bibnamefont
  {Walker}}, \bibinfo {author} {\bibfnamefont {L.}~\bibnamefont {Kolanz}},
  \bibinfo {author} {\bibfnamefont {J.}~\bibnamefont {Venzke}},\ and\ \bibinfo
  {author} {\bibfnamefont {A.}~\bibnamefont {Becker}},\ }\bibfield  {title}
  {\bibinfo {title} {Enhanced ionization of counter-rotating electrons via
  doorway states in ultrashort circularly polarized laser pulses},\ }\href
  {https://doi.org/10.1103/PhysRevA.103.L061101} {\bibfield  {journal}
  {\bibinfo  {journal} {Phys. Rev. A}\ }\textbf {\bibinfo {volume} {103}},\
  \bibinfo {pages} {L061101} (\bibinfo {year} {2021})}\BibitemShut {NoStop}%
\bibitem [{\citenamefont {Klaiber}\ \emph {et~al.}(2014)\citenamefont
  {Klaiber}, \citenamefont {Yakaboylu}, \citenamefont {M{\"u}ller},
  \citenamefont {Bauke}, \citenamefont {Paulus},\ and\ \citenamefont
  {Hatsagortsyan}}]{Klaiber_2014}%
  \BibitemOpen
  \bibfield  {author} {\bibinfo {author} {\bibfnamefont {M.}~\bibnamefont
  {Klaiber}}, \bibinfo {author} {\bibfnamefont {E.}~\bibnamefont {Yakaboylu}},
  \bibinfo {author} {\bibfnamefont {C.}~\bibnamefont {M{\"u}ller}}, \bibinfo
  {author} {\bibfnamefont {H.}~\bibnamefont {Bauke}}, \bibinfo {author}
  {\bibfnamefont {G.~G.}\ \bibnamefont {Paulus}},\ and\ \bibinfo {author}
  {\bibfnamefont {K.~Z.}\ \bibnamefont {Hatsagortsyan}},\ }\bibfield  {title}
  {\bibinfo {title} {Spin dynamics in relativistic ionization with highly
  charged ions in super-strong laser fields},\ }\href@noop {} {\bibfield
  {journal} {\bibinfo  {journal} {Journal of Physics B: Atomic, Molecular and
  Optical Physics}\ }\textbf {\bibinfo {volume} {47}},\ \bibinfo {pages}
  {065603} (\bibinfo {year} {2014})}\BibitemShut {NoStop}%
\bibitem [{\citenamefont {Yakaboylu}\ \emph {et~al.}(2015)\citenamefont
  {Yakaboylu}, \citenamefont {Klaiber},\ and\ \citenamefont
  {Hatsagortsyan}}]{Klaiber_2015}%
  \BibitemOpen
  \bibfield  {author} {\bibinfo {author} {\bibfnamefont {E.}~\bibnamefont
  {Yakaboylu}}, \bibinfo {author} {\bibfnamefont {M.}~\bibnamefont {Klaiber}},\
  and\ \bibinfo {author} {\bibfnamefont {K.~Z.}\ \bibnamefont
  {Hatsagortsyan}},\ }\bibfield  {title} {\bibinfo {title} {Above-threshold
  ionization with highly charged ions in superstrong laser fields. iii. spin
  effects and their dependence on laser polarization},\ }\href
  {https://doi.org/10.1103/PhysRevA.91.063407} {\bibfield  {journal} {\bibinfo
  {journal} {Phys. Rev. A}\ }\textbf {\bibinfo {volume} {91}},\ \bibinfo
  {pages} {063407} (\bibinfo {year} {2015})}\BibitemShut {NoStop}%
\bibitem [{\citenamefont {Corkum}(1993)}]{corkum1993plasma}%
  \BibitemOpen
  \bibfield  {author} {\bibinfo {author} {\bibfnamefont {P.~B.}\ \bibnamefont
  {Corkum}},\ }\bibfield  {title} {\bibinfo {title} {Plasma perspective on
  strong field multiphoton ionization},\ }\href
  {https://doi.org/10.1103/PhysRevLett.71.1994} {\bibfield  {journal} {\bibinfo
   {journal} {Phys. Rev. Lett.}\ }\textbf {\bibinfo {volume} {71}},\ \bibinfo
  {pages} {1994} (\bibinfo {year} {1993})}\BibitemShut {NoStop}%
\bibitem [{\citenamefont {Zille}\ \emph {et~al.}(2017)\citenamefont {Zille},
  \citenamefont {Seipt}, \citenamefont {Moller}, \citenamefont {Fritzsche},
  \citenamefont {Paulus},\ and\ \citenamefont {Milosevic}}]{Zille_2017}%
  \BibitemOpen
  \bibfield  {author} {\bibinfo {author} {\bibfnamefont {D.}~\bibnamefont
  {Zille}}, \bibinfo {author} {\bibfnamefont {D.}~\bibnamefont {Seipt}},
  \bibinfo {author} {\bibfnamefont {M.}~\bibnamefont {Moller}}, \bibinfo
  {author} {\bibfnamefont {S.}~\bibnamefont {Fritzsche}}, \bibinfo {author}
  {\bibfnamefont {G.~G.}\ \bibnamefont {Paulus}},\ and\ \bibinfo {author}
  {\bibfnamefont {D.~B.}\ \bibnamefont {Milosevic}},\ }\bibfield  {title}
  {\bibinfo {title} {{Spin-dependent quantum theory of high-order
  above-threshold ionization}},\ }\href@noop {} {\bibfield  {journal} {\bibinfo
   {journal} {Phys. Rev. A}\ }\textbf {\bibinfo {volume} {95}},\ \bibinfo
  {pages} {063408} (\bibinfo {year} {2017})}\BibitemShut {NoStop}%
\bibitem [{\citenamefont {Walser}\ \emph {et~al.}(1999)\citenamefont {Walser},
  \citenamefont {Szymanowski},\ and\ \citenamefont {Keitel}}]{Walser_1999}%
  \BibitemOpen
  \bibfield  {author} {\bibinfo {author} {\bibfnamefont {M.~W.}\ \bibnamefont
  {Walser}}, \bibinfo {author} {\bibfnamefont {C.}~\bibnamefont
  {Szymanowski}},\ and\ \bibinfo {author} {\bibfnamefont {C.~H.}\ \bibnamefont
  {Keitel}},\ }\bibfield  {title} {\bibinfo {title} {{Influence of spin-laser
  interaction on relativistic harmonic generation}},\ }\href@noop {} {\bibfield
   {journal} {\bibinfo  {journal} {Europhysics Letters}\ }\textbf {\bibinfo
  {volume} {48}},\ \bibinfo {pages} {533} (\bibinfo {year} {1999})}\BibitemShut
  {NoStop}%
\bibitem [{\citenamefont {Hu}\ and\ \citenamefont {Keitel}(1999)}]{Hu_1999}%
  \BibitemOpen
  \bibfield  {author} {\bibinfo {author} {\bibfnamefont {S.~X.}\ \bibnamefont
  {Hu}}\ and\ \bibinfo {author} {\bibfnamefont {C.~H.}\ \bibnamefont
  {Keitel}},\ }\bibfield  {title} {\bibinfo {title} {Spin signatures in intense
  laser-ion interaction},\ }\href {https://doi.org/10.1103/PhysRevLett.83.4709}
  {\bibfield  {journal} {\bibinfo  {journal} {Phys. Rev. Lett.}\ }\textbf
  {\bibinfo {volume} {83}},\ \bibinfo {pages} {4709} (\bibinfo {year}
  {1999})}\BibitemShut {NoStop}%
\bibitem [{\citenamefont {Walser}\ \emph {et~al.}(2002)\citenamefont {Walser},
  \citenamefont {Urbach}, \citenamefont {Hatsagortsyan}, \citenamefont {Hu},\
  and\ \citenamefont {Keitel}}]{Walser_2002}%
  \BibitemOpen
  \bibfield  {author} {\bibinfo {author} {\bibfnamefont {M.~W.}\ \bibnamefont
  {Walser}}, \bibinfo {author} {\bibfnamefont {D.~J.}\ \bibnamefont {Urbach}},
  \bibinfo {author} {\bibfnamefont {K.~Z.}\ \bibnamefont {Hatsagortsyan}},
  \bibinfo {author} {\bibfnamefont {S.~X.}\ \bibnamefont {Hu}},\ and\ \bibinfo
  {author} {\bibfnamefont {C.~H.}\ \bibnamefont {Keitel}},\ }\bibfield  {title}
  {\bibinfo {title} {Spin and radiation in intense laser fields},\ }\href
  {https://doi.org/10.1103/PhysRevA.65.043410} {\bibfield  {journal} {\bibinfo
  {journal} {Phys. Rev. A}\ }\textbf {\bibinfo {volume} {65}},\ \bibinfo
  {pages} {043410} (\bibinfo {year} {2002})}\BibitemShut {NoStop}%
\bibitem [{\citenamefont {Maxwell}\ and\ \citenamefont
  {Madsen}(2023)}]{maxwell2023relativistic}%
  \BibitemOpen
  \bibfield  {author} {\bibinfo {author} {\bibfnamefont {A.~S.}\ \bibnamefont
  {Maxwell}}\ and\ \bibinfo {author} {\bibfnamefont {L.~B.}\ \bibnamefont
  {Madsen}},\ }\bibfield  {title} {\bibinfo {title} {Relativistic and
  spin-orbit dynamics at non-relativistic intensities in strong-field
  ionization},\ }\href@noop {} {\bibfield  {journal} {\bibinfo  {journal}
  {arXiv preprint arXiv:2308.15374}\ } (\bibinfo {year} {2023})}\BibitemShut
  {NoStop}%
\bibitem [{\citenamefont {Rohringer}\ and\ \citenamefont
  {Santra}(2009)}]{rohringer2009multichannel}%
  \BibitemOpen
  \bibfield  {author} {\bibinfo {author} {\bibfnamefont {N.}~\bibnamefont
  {Rohringer}}\ and\ \bibinfo {author} {\bibfnamefont {R.}~\bibnamefont
  {Santra}},\ }\bibfield  {title} {\bibinfo {title} {Multichannel coherence in
  strong-field ionization},\ }\href@noop {} {\bibfield  {journal} {\bibinfo
  {journal} {Physical Review A}\ }\textbf {\bibinfo {volume} {79}},\ \bibinfo
  {pages} {053402} (\bibinfo {year} {2009})}\BibitemShut {NoStop}%
\bibitem [{\citenamefont {Barth}\ and\ \citenamefont
  {Smirnova}(2014{\natexlab{b}})}]{barth2014hole}%
  \BibitemOpen
  \bibfield  {author} {\bibinfo {author} {\bibfnamefont {I.}~\bibnamefont
  {Barth}}\ and\ \bibinfo {author} {\bibfnamefont {O.}~\bibnamefont
  {Smirnova}},\ }\bibfield  {title} {\bibinfo {title} {Hole dynamics and spin
  currents after ionization in strong circularly polarized laser fields},\
  }\href@noop {} {\bibfield  {journal} {\bibinfo  {journal} {Journal of Physics
  B: Atomic, Molecular and Optical Physics}\ }\textbf {\bibinfo {volume}
  {47}},\ \bibinfo {pages} {204020} (\bibinfo {year}
  {2014}{\natexlab{b}})}\BibitemShut {NoStop}%
\bibitem [{\citenamefont {K{\"u}bel}\ \emph {et~al.}(2019)\citenamefont
  {K{\"u}bel}, \citenamefont {Dube}, \citenamefont {Naumov}, \citenamefont
  {Villeneuve}, \citenamefont {Corkum},\ and\ \citenamefont
  {Staudte}}]{kubel2019spatiotemporal}%
  \BibitemOpen
  \bibfield  {author} {\bibinfo {author} {\bibfnamefont {M.}~\bibnamefont
  {K{\"u}bel}}, \bibinfo {author} {\bibfnamefont {Z.}~\bibnamefont {Dube}},
  \bibinfo {author} {\bibfnamefont {A.~Y.}\ \bibnamefont {Naumov}}, \bibinfo
  {author} {\bibfnamefont {D.~M.}\ \bibnamefont {Villeneuve}}, \bibinfo
  {author} {\bibfnamefont {P.~B.}\ \bibnamefont {Corkum}},\ and\ \bibinfo
  {author} {\bibfnamefont {A.}~\bibnamefont {Staudte}},\ }\bibfield  {title}
  {\bibinfo {title} {Spatiotemporal imaging of valence electron motion},\
  }\href@noop {} {\bibfield  {journal} {\bibinfo  {journal} {Nature
  Communications}\ }\textbf {\bibinfo {volume} {10}},\ \bibinfo {pages} {1042}
  (\bibinfo {year} {2019})}\BibitemShut {NoStop}%
\bibitem [{\citenamefont {Mayer}\ \emph {et~al.}(2022)\citenamefont {Mayer},
  \citenamefont {Beaulieu}, \citenamefont {Jim{\'e}nez-Gal{\'a}n},
  \citenamefont {Patchkovskii}, \citenamefont {Kornilov}, \citenamefont
  {Descamps}, \citenamefont {Petit}, \citenamefont {Smirnova}, \citenamefont
  {Mairesse},\ and\ \citenamefont {Ivanov}}]{mayer2022role}%
  \BibitemOpen
  \bibfield  {author} {\bibinfo {author} {\bibfnamefont {N.}~\bibnamefont
  {Mayer}}, \bibinfo {author} {\bibfnamefont {S.}~\bibnamefont {Beaulieu}},
  \bibinfo {author} {\bibfnamefont {{\'A}.}~\bibnamefont
  {Jim{\'e}nez-Gal{\'a}n}}, \bibinfo {author} {\bibfnamefont {S.}~\bibnamefont
  {Patchkovskii}}, \bibinfo {author} {\bibfnamefont {O.}~\bibnamefont
  {Kornilov}}, \bibinfo {author} {\bibfnamefont {D.}~\bibnamefont {Descamps}},
  \bibinfo {author} {\bibfnamefont {S.}~\bibnamefont {Petit}}, \bibinfo
  {author} {\bibfnamefont {O.}~\bibnamefont {Smirnova}}, \bibinfo {author}
  {\bibfnamefont {Y.}~\bibnamefont {Mairesse}},\ and\ \bibinfo {author}
  {\bibfnamefont {M.}~\bibnamefont {Ivanov}},\ }\bibfield  {title} {\bibinfo
  {title} {Role of spin-orbit coupling in high-order harmonic generation
  revealed by supercycle rydberg trajectories},\ }\href@noop {} {\bibfield
  {journal} {\bibinfo  {journal} {Physical Review Letters}\ }\textbf {\bibinfo
  {volume} {129}},\ \bibinfo {pages} {173202} (\bibinfo {year}
  {2022})}\BibitemShut {NoStop}%
\bibitem [{\citenamefont {Stewart}\ \emph {et~al.}(2023)\citenamefont
  {Stewart}, \citenamefont {Hoerner}, \citenamefont {Debrah}, \citenamefont
  {Lee}, \citenamefont {Schlegel},\ and\ \citenamefont
  {Li}}]{stewart2023attosecond}%
  \BibitemOpen
  \bibfield  {author} {\bibinfo {author} {\bibfnamefont {G.~A.}\ \bibnamefont
  {Stewart}}, \bibinfo {author} {\bibfnamefont {P.}~\bibnamefont {Hoerner}},
  \bibinfo {author} {\bibfnamefont {D.~A.}\ \bibnamefont {Debrah}}, \bibinfo
  {author} {\bibfnamefont {S.~K.}\ \bibnamefont {Lee}}, \bibinfo {author}
  {\bibfnamefont {H.~B.}\ \bibnamefont {Schlegel}},\ and\ \bibinfo {author}
  {\bibfnamefont {W.}~\bibnamefont {Li}},\ }\bibfield  {title} {\bibinfo
  {title} {Attosecond imaging of electronic wave packets},\ }\href@noop {}
  {\bibfield  {journal} {\bibinfo  {journal} {Physical Review Letters}\
  }\textbf {\bibinfo {volume} {130}},\ \bibinfo {pages} {083202} (\bibinfo
  {year} {2023})}\BibitemShut {NoStop}%
\bibitem [{\citenamefont {Carlstr{\"o}m}\ \emph
  {et~al.}(2022{\natexlab{a}})\citenamefont {Carlstr{\"o}m}, \citenamefont
  {Spanner},\ and\ \citenamefont {Patchkovskii}}]{carlstrom2022general}%
  \BibitemOpen
  \bibfield  {author} {\bibinfo {author} {\bibfnamefont {S.}~\bibnamefont
  {Carlstr{\"o}m}}, \bibinfo {author} {\bibfnamefont {M.}~\bibnamefont
  {Spanner}},\ and\ \bibinfo {author} {\bibfnamefont {S.}~\bibnamefont
  {Patchkovskii}},\ }\bibfield  {title} {\bibinfo {title} {General
  time-dependent configuration-interaction singles. i. molecular case},\
  }\href@noop {} {\bibfield  {journal} {\bibinfo  {journal} {Physical Review
  A}\ }\textbf {\bibinfo {volume} {106}},\ \bibinfo {pages} {043104} (\bibinfo
  {year} {2022}{\natexlab{a}})}\BibitemShut {NoStop}%
\bibitem [{\citenamefont {Carlstr{\"o}m}\ \emph
  {et~al.}(2022{\natexlab{b}})\citenamefont {Carlstr{\"o}m}, \citenamefont
  {Bertolino}, \citenamefont {Dahlstr{\"o}m},\ and\ \citenamefont
  {Patchkovskii}}]{carlstrom2022general2}%
  \BibitemOpen
  \bibfield  {author} {\bibinfo {author} {\bibfnamefont {S.}~\bibnamefont
  {Carlstr{\"o}m}}, \bibinfo {author} {\bibfnamefont {M.}~\bibnamefont
  {Bertolino}}, \bibinfo {author} {\bibfnamefont {J.~M.}\ \bibnamefont
  {Dahlstr{\"o}m}},\ and\ \bibinfo {author} {\bibfnamefont {S.}~\bibnamefont
  {Patchkovskii}},\ }\bibfield  {title} {\bibinfo {title} {General
  time-dependent configuration-interaction singles. ii. atomic case},\
  }\href@noop {} {\bibfield  {journal} {\bibinfo  {journal} {Physical Review
  A}\ }\textbf {\bibinfo {volume} {106}},\ \bibinfo {pages} {042806} (\bibinfo
  {year} {2022}{\natexlab{b}})}\BibitemShut {NoStop}%
\bibitem [{\citenamefont {Carlstr\"om}\ \emph {et~al.}(2023)\citenamefont
  {Carlstr\"om}, \citenamefont {Dahlstr\"om}, \citenamefont {Ivanov},
  \citenamefont {Smirnova},\ and\ \citenamefont
  {Patchkovskii}}]{carlstrom2023control}%
  \BibitemOpen
  \bibfield  {author} {\bibinfo {author} {\bibfnamefont {S.}~\bibnamefont
  {Carlstr\"om}}, \bibinfo {author} {\bibfnamefont {J.~M.}\ \bibnamefont
  {Dahlstr\"om}}, \bibinfo {author} {\bibfnamefont {M.~Y.}\ \bibnamefont
  {Ivanov}}, \bibinfo {author} {\bibfnamefont {O.}~\bibnamefont {Smirnova}},\
  and\ \bibinfo {author} {\bibfnamefont {S.}~\bibnamefont {Patchkovskii}},\
  }\bibfield  {title} {\bibinfo {title} {Control of spin polarization through
  recollisions},\ }\href {https://doi.org/10.1103/PhysRevA.108.043104}
  {\bibfield  {journal} {\bibinfo  {journal} {Phys. Rev. A}\ }\textbf {\bibinfo
  {volume} {108}},\ \bibinfo {pages} {043104} (\bibinfo {year}
  {2023})}\BibitemShut {NoStop}%
\bibitem [{\citenamefont {Keldysh}(1965)}]{keldysh1965ionization}%
  \BibitemOpen
  \bibfield  {author} {\bibinfo {author} {\bibfnamefont {L.}~\bibnamefont
  {Keldysh}},\ }\bibfield  {title} {\bibinfo {title} {Ionization in the field
  of a strong electromagnetic wave},\ }\href@noop {} {\bibfield  {journal}
  {\bibinfo  {journal} {Sov. Phys. JETP}\ }\textbf {\bibinfo {volume} {20}},\
  \bibinfo {pages} {1307} (\bibinfo {year} {1965})}\BibitemShut {NoStop}%
\bibitem [{\citenamefont {Faisal}(1973)}]{faisal1973multiple}%
  \BibitemOpen
  \bibfield  {author} {\bibinfo {author} {\bibfnamefont {F.~H.~M.}\
  \bibnamefont {Faisal}},\ }\bibfield  {title} {\bibinfo {title} {Multiple
  absorption of laser photons by atoms},\ }\href@noop {} {\bibfield  {journal}
  {\bibinfo  {journal} {J. Phys. B.}\ }\textbf {\bibinfo {volume} {6}},\
  \bibinfo {pages} {L89} (\bibinfo {year} {1973})}\BibitemShut {NoStop}%
\bibitem [{\citenamefont {Reiss}(1980)}]{reiss1980effect}%
  \BibitemOpen
  \bibfield  {author} {\bibinfo {author} {\bibfnamefont {H.~R.}\ \bibnamefont
  {Reiss}},\ }\bibfield  {title} {\bibinfo {title} {Effect of an intense
  electromagnetic field on a weakly bound system},\ }\href
  {https://doi.org/10.1103/PhysRevA.22.1786} {\bibfield  {journal} {\bibinfo
  {journal} {Phys. Rev. A}\ }\textbf {\bibinfo {volume} {22}},\ \bibinfo
  {pages} {1786} (\bibinfo {year} {1980})}\BibitemShut {NoStop}%
\bibitem [{sup()}]{supp}%
  \BibitemOpen
  \href@noop {} {}\bibinfo {howpublished} {See the Supplemental Materials for
  the details.}\BibitemShut {Stop}%
\bibitem [{\citenamefont {Becker}\ \emph {et~al.}(2002)\citenamefont {Becker},
  \citenamefont {Grasbon}, \citenamefont {Kopold}, \citenamefont
  {Milo{\v{s}}evi{\'c}}, \citenamefont {Paulus},\ and\ \citenamefont
  {Walther}}]{becker2002above}%
  \BibitemOpen
  \bibfield  {author} {\bibinfo {author} {\bibfnamefont {W.}~\bibnamefont
  {Becker}}, \bibinfo {author} {\bibfnamefont {F.}~\bibnamefont {Grasbon}},
  \bibinfo {author} {\bibfnamefont {R.}~\bibnamefont {Kopold}}, \bibinfo
  {author} {\bibfnamefont {D.}~\bibnamefont {Milo{\v{s}}evi{\'c}}}, \bibinfo
  {author} {\bibfnamefont {G.}~\bibnamefont {Paulus}},\ and\ \bibinfo {author}
  {\bibfnamefont {H.}~\bibnamefont {Walther}},\ }\bibfield  {title} {\bibinfo
  {title} {Above-threshold ionization: From classical features to quantum
  effects},\ }\href@noop {} {\bibfield  {journal} {\bibinfo  {journal}
  {Advances in Atomic, molecular, and optical physics}\ }\textbf {\bibinfo
  {volume} {48}},\ \bibinfo {pages} {35} (\bibinfo {year} {2002})}\BibitemShut
  {NoStop}%
\bibitem [{\citenamefont {Sakurai}\ and\ \citenamefont
  {Napolitano}(2020)}]{sakurai2020modern}%
  \BibitemOpen
  \bibfield  {author} {\bibinfo {author} {\bibfnamefont {J.}~\bibnamefont
  {Sakurai}}\ and\ \bibinfo {author} {\bibfnamefont {J.}~\bibnamefont
  {Napolitano}},\ }\href {https://books.google.de/books?id=XwpuzQEACAAJ} {\emph
  {\bibinfo {title} {Modern Quantum Mechanics}}}\ (\bibinfo  {publisher}
  {Cambridge University Press},\ \bibinfo {year} {2020})\BibitemShut {NoStop}%
\bibitem [{\citenamefont {Bargmann}\ \emph {et~al.}(1959)\citenamefont
  {Bargmann}, \citenamefont {Michel},\ and\ \citenamefont
  {Telegdi}}]{bargmann1959precession}%
  \BibitemOpen
  \bibfield  {author} {\bibinfo {author} {\bibfnamefont {V.}~\bibnamefont
  {Bargmann}}, \bibinfo {author} {\bibfnamefont {L.}~\bibnamefont {Michel}},\
  and\ \bibinfo {author} {\bibfnamefont {V.}~\bibnamefont {Telegdi}},\
  }\bibfield  {title} {\bibinfo {title} {Precession of the polarization of
  particles moving in a homogeneous electromagnetic field},\ }\href@noop {}
  {\bibfield  {journal} {\bibinfo  {journal} {Physical Review Letters}\
  }\textbf {\bibinfo {volume} {2}},\ \bibinfo {pages} {435} (\bibinfo {year}
  {1959})}\BibitemShut {NoStop}%
\bibitem [{\citenamefont {Perelomov}\ \emph {et~al.}(1966)\citenamefont
  {Perelomov}, \citenamefont {Popov},\ and\ \citenamefont
  {Terent'ev}}]{perelomov1966ionization}%
  \BibitemOpen
  \bibfield  {author} {\bibinfo {author} {\bibfnamefont {A.~M.}\ \bibnamefont
  {Perelomov}}, \bibinfo {author} {\bibfnamefont {V.~S.}\ \bibnamefont
  {Popov}},\ and\ \bibinfo {author} {\bibfnamefont {M.~V.}\ \bibnamefont
  {Terent'ev}},\ }\bibfield  {title} {\bibinfo {title} {Ionization of atoms in
  an alternating electric field},\ }\href@noop {} {\bibfield  {journal}
  {\bibinfo  {journal} {Sov. Phys. JETP}\ }\textbf {\bibinfo {volume} {23}},\
  \bibinfo {pages} {924} (\bibinfo {year} {1966})}\BibitemShut {NoStop}%
\bibitem [{\citenamefont {Delone}\ and\ \citenamefont
  {Krainov}(1991)}]{delone1991energy}%
  \BibitemOpen
  \bibfield  {author} {\bibinfo {author} {\bibfnamefont {N.~B.}\ \bibnamefont
  {Delone}}\ and\ \bibinfo {author} {\bibfnamefont {V.~P.}\ \bibnamefont
  {Krainov}},\ }\bibfield  {title} {\bibinfo {title} {Energy and angular
  electron spectra for the tunnel ionization of atoms by strong low-frequency
  radiation},\ }\href@noop {} {\bibfield  {journal} {\bibinfo  {journal} {J.
  Opt. Soc. Am. B}\ }\textbf {\bibinfo {volume} {8}},\ \bibinfo {pages} {1207}
  (\bibinfo {year} {1991})}\BibitemShut {NoStop}%
\bibitem [{\citenamefont {Keil}\ \emph {et~al.}(2016)\citenamefont {Keil},
  \citenamefont {Popruzhenko},\ and\ \citenamefont {Bauer}}]{keil2016laser}%
  \BibitemOpen
  \bibfield  {author} {\bibinfo {author} {\bibfnamefont {T.}~\bibnamefont
  {Keil}}, \bibinfo {author} {\bibfnamefont {S.~V.}\ \bibnamefont
  {Popruzhenko}},\ and\ \bibinfo {author} {\bibfnamefont {D.}~\bibnamefont
  {Bauer}},\ }\bibfield  {title} {\bibinfo {title} {Laser-driven recollisions
  under the coulomb barrier},\ }\href
  {https://doi.org/10.1103/PhysRevLett.117.243003} {\bibfield  {journal}
  {\bibinfo  {journal} {Phys. Rev. Lett.}\ }\textbf {\bibinfo {volume} {117}},\
  \bibinfo {pages} {243003} (\bibinfo {year} {2016})}\BibitemShut {NoStop}%
\bibitem [{\citenamefont {He}\ \emph {et~al.}(2018)\citenamefont {He},
  \citenamefont {Klaiber}, \citenamefont {Hatsagortsyan},\ and\ \citenamefont
  {Keitel}}]{he2018high}%
  \BibitemOpen
  \bibfield  {author} {\bibinfo {author} {\bibfnamefont {P.-L.}\ \bibnamefont
  {He}}, \bibinfo {author} {\bibfnamefont {M.}~\bibnamefont {Klaiber}},
  \bibinfo {author} {\bibfnamefont {K.~Z.}\ \bibnamefont {Hatsagortsyan}},\
  and\ \bibinfo {author} {\bibfnamefont {C.~H.}\ \bibnamefont {Keitel}},\
  }\bibfield  {title} {\bibinfo {title} {High-energy direct photoelectron
  spectroscopy in strong-field ionization},\ }\href@noop {} {\bibfield
  {journal} {\bibinfo  {journal} {Physical Review A}\ }\textbf {\bibinfo
  {volume} {98}},\ \bibinfo {pages} {053428} (\bibinfo {year}
  {2018})}\BibitemShut {NoStop}%
\bibitem [{\citenamefont {Li}\ \emph {et~al.}(2014)\citenamefont {Li},
  \citenamefont {Geng}, \citenamefont {Liu}, \citenamefont {Deng},
  \citenamefont {Wu}, \citenamefont {Peng}, \citenamefont {Gong},\ and\
  \citenamefont {Liu}}]{li2014classical}%
  \BibitemOpen
  \bibfield  {author} {\bibinfo {author} {\bibfnamefont {M.}~\bibnamefont
  {Li}}, \bibinfo {author} {\bibfnamefont {J.-W.}\ \bibnamefont {Geng}},
  \bibinfo {author} {\bibfnamefont {H.}~\bibnamefont {Liu}}, \bibinfo {author}
  {\bibfnamefont {Y.}~\bibnamefont {Deng}}, \bibinfo {author} {\bibfnamefont
  {C.}~\bibnamefont {Wu}}, \bibinfo {author} {\bibfnamefont {L.-Y.}\
  \bibnamefont {Peng}}, \bibinfo {author} {\bibfnamefont {Q.}~\bibnamefont
  {Gong}},\ and\ \bibinfo {author} {\bibfnamefont {Y.}~\bibnamefont {Liu}},\
  }\bibfield  {title} {\bibinfo {title} {Classical-quantum correspondence for
  above-threshold ionization},\ }\href@noop {} {\bibfield  {journal} {\bibinfo
  {journal} {Physical review letters}\ }\textbf {\bibinfo {volume} {112}},\
  \bibinfo {pages} {113002} (\bibinfo {year} {2014})}\BibitemShut {NoStop}%
\bibitem [{\citenamefont {Lai}\ \emph {et~al.}(2015)\citenamefont {Lai},
  \citenamefont {Poli}, \citenamefont {Schomerus},\ and\ \citenamefont
  {Faria}}]{PhysRevA.92.043407}%
  \BibitemOpen
  \bibfield  {author} {\bibinfo {author} {\bibfnamefont {X.-Y.}\ \bibnamefont
  {Lai}}, \bibinfo {author} {\bibfnamefont {C.}~\bibnamefont {Poli}}, \bibinfo
  {author} {\bibfnamefont {H.}~\bibnamefont {Schomerus}},\ and\ \bibinfo
  {author} {\bibfnamefont {C.~F. d.~M.}\ \bibnamefont {Faria}},\ }\bibfield
  {title} {\bibinfo {title} {Influence of the coulomb potential on
  above-threshold ionization: A quantum-orbit analysis beyond the strong-field
  approximation},\ }\href {https://doi.org/10.1103/PhysRevA.92.043407}
  {\bibfield  {journal} {\bibinfo  {journal} {Phys. Rev. A}\ }\textbf {\bibinfo
  {volume} {92}},\ \bibinfo {pages} {043407} (\bibinfo {year}
  {2015})}\BibitemShut {NoStop}%
\bibitem [{\citenamefont {Shvetsov-Shilovski}\ and\ \citenamefont
  {Lein}(2019)}]{PhysRevA.100.053411}%
  \BibitemOpen
  \bibfield  {author} {\bibinfo {author} {\bibfnamefont {N.~I.}\ \bibnamefont
  {Shvetsov-Shilovski}}\ and\ \bibinfo {author} {\bibfnamefont
  {M.}~\bibnamefont {Lein}},\ }\bibfield  {title} {\bibinfo {title}
  {Semiclassical two-step model with quantum input: Quantum-classical approach
  to strong-field ionization},\ }\href
  {https://doi.org/10.1103/PhysRevA.100.053411} {\bibfield  {journal} {\bibinfo
   {journal} {Phys. Rev. A}\ }\textbf {\bibinfo {volume} {100}},\ \bibinfo
  {pages} {053411} (\bibinfo {year} {2019})}\BibitemShut {NoStop}%
\bibitem [{\citenamefont {He}\ \emph {et~al.}(2015)\citenamefont {He},
  \citenamefont {Takemoto},\ and\ \citenamefont {He}}]{he2015photoelectron}%
  \BibitemOpen
  \bibfield  {author} {\bibinfo {author} {\bibfnamefont {P.-L.}\ \bibnamefont
  {He}}, \bibinfo {author} {\bibfnamefont {N.}~\bibnamefont {Takemoto}},\ and\
  \bibinfo {author} {\bibfnamefont {F.}~\bibnamefont {He}},\ }\bibfield
  {title} {\bibinfo {title} {Photoelectron momentum distributions of atomic and
  molecular systems in strong circularly or elliptically polarized laser
  fields},\ }\href@noop {} {\bibfield  {journal} {\bibinfo  {journal} {Physical
  Review A}\ }\textbf {\bibinfo {volume} {91}},\ \bibinfo {pages} {063413}
  (\bibinfo {year} {2015})}\BibitemShut {NoStop}%
\bibitem [{\citenamefont {Huismans}\ \emph {et~al.}(2011)\citenamefont
  {Huismans}, \citenamefont {Rouz{\'e}e}, \citenamefont {Gijsbertsen},
  \citenamefont {Jungmann}, \citenamefont {Smolkowska}, \citenamefont {Logman},
  \citenamefont {Lepine}, \citenamefont {Cauchy}, \citenamefont {Zamith},
  \citenamefont {Marchenko} \emph {et~al.}}]{huismans2011time}%
  \BibitemOpen
  \bibfield  {author} {\bibinfo {author} {\bibfnamefont {Y.}~\bibnamefont
  {Huismans}}, \bibinfo {author} {\bibfnamefont {A.}~\bibnamefont
  {Rouz{\'e}e}}, \bibinfo {author} {\bibfnamefont {A.}~\bibnamefont
  {Gijsbertsen}}, \bibinfo {author} {\bibfnamefont {J.}~\bibnamefont
  {Jungmann}}, \bibinfo {author} {\bibfnamefont {A.}~\bibnamefont
  {Smolkowska}}, \bibinfo {author} {\bibfnamefont {P.}~\bibnamefont {Logman}},
  \bibinfo {author} {\bibfnamefont {F.}~\bibnamefont {Lepine}}, \bibinfo
  {author} {\bibfnamefont {C.}~\bibnamefont {Cauchy}}, \bibinfo {author}
  {\bibfnamefont {S.}~\bibnamefont {Zamith}}, \bibinfo {author} {\bibfnamefont
  {T.}~\bibnamefont {Marchenko}}, \emph {et~al.},\ }\bibfield  {title}
  {\bibinfo {title} {Time-resolved holography with photoelectrons},\
  }\href@noop {} {\bibfield  {journal} {\bibinfo  {journal} {Science}\ }\textbf
  {\bibinfo {volume} {331}},\ \bibinfo {pages} {61} (\bibinfo {year}
  {2011})}\BibitemShut {NoStop}%
\bibitem [{\citenamefont {Huismans}\ \emph {et~al.}(2012)\citenamefont
  {Huismans}, \citenamefont {Gijsbertsen}, \citenamefont {Smolkowska},
  \citenamefont {Jungmann}, \citenamefont {Rouz\'ee}, \citenamefont {Logman},
  \citenamefont {L\'epine}, \citenamefont {Cauchy}, \citenamefont {Zamith},
  \citenamefont {Marchenko}, \citenamefont {Bakker}, \citenamefont {Berden},
  \citenamefont {Redlich}, \citenamefont {van~der Meer}, \citenamefont
  {Ivanov}, \citenamefont {Yan}, \citenamefont {Bauer}, \citenamefont
  {Smirnova},\ and\ \citenamefont {Vrakking}}]{Huismans_2012}%
  \BibitemOpen
  \bibfield  {author} {\bibinfo {author} {\bibfnamefont {Y.}~\bibnamefont
  {Huismans}}, \bibinfo {author} {\bibfnamefont {A.}~\bibnamefont
  {Gijsbertsen}}, \bibinfo {author} {\bibfnamefont {A.~S.}\ \bibnamefont
  {Smolkowska}}, \bibinfo {author} {\bibfnamefont {J.~H.}\ \bibnamefont
  {Jungmann}}, \bibinfo {author} {\bibfnamefont {A.}~\bibnamefont {Rouz\'ee}},
  \bibinfo {author} {\bibfnamefont {P.~S. W.~M.}\ \bibnamefont {Logman}},
  \bibinfo {author} {\bibfnamefont {F.}~\bibnamefont {L\'epine}}, \bibinfo
  {author} {\bibfnamefont {C.}~\bibnamefont {Cauchy}}, \bibinfo {author}
  {\bibfnamefont {S.}~\bibnamefont {Zamith}}, \bibinfo {author} {\bibfnamefont
  {T.}~\bibnamefont {Marchenko}}, \bibinfo {author} {\bibfnamefont {J.~M.}\
  \bibnamefont {Bakker}}, \bibinfo {author} {\bibfnamefont {G.}~\bibnamefont
  {Berden}}, \bibinfo {author} {\bibfnamefont {B.}~\bibnamefont {Redlich}},
  \bibinfo {author} {\bibfnamefont {A.~F.~G.}\ \bibnamefont {van~der Meer}},
  \bibinfo {author} {\bibfnamefont {M.~Y.}\ \bibnamefont {Ivanov}}, \bibinfo
  {author} {\bibfnamefont {T.-M.}\ \bibnamefont {Yan}}, \bibinfo {author}
  {\bibfnamefont {D.}~\bibnamefont {Bauer}}, \bibinfo {author} {\bibfnamefont
  {O.}~\bibnamefont {Smirnova}},\ and\ \bibinfo {author} {\bibfnamefont
  {M.~J.~J.}\ \bibnamefont {Vrakking}},\ }\bibfield  {title} {\bibinfo {title}
  {Scaling laws for photoelectron holography in the midinfrared wavelength
  regime},\ }\href {https://doi.org/10.1103/PhysRevLett.109.013002} {\bibfield
  {journal} {\bibinfo  {journal} {Phys. Rev. Lett.}\ }\textbf {\bibinfo
  {volume} {109}},\ \bibinfo {pages} {013002} (\bibinfo {year}
  {2012})}\BibitemShut {NoStop}%
\bibitem [{\citenamefont {Bian}\ and\ \citenamefont
  {Bandrauk}(2012)}]{bian2012attosecond}%
  \BibitemOpen
  \bibfield  {author} {\bibinfo {author} {\bibfnamefont {X.-B.}\ \bibnamefont
  {Bian}}\ and\ \bibinfo {author} {\bibfnamefont {A.~D.}\ \bibnamefont
  {Bandrauk}},\ }\bibfield  {title} {\bibinfo {title} {Attosecond time-resolved
  imaging of molecular structure by photoelectron holography},\ }\href@noop {}
  {\bibfield  {journal} {\bibinfo  {journal} {Physical Review Letters}\
  }\textbf {\bibinfo {volume} {108}},\ \bibinfo {pages} {263003} (\bibinfo
  {year} {2012})}\BibitemShut {NoStop}%
\bibitem [{\citenamefont {Zhou}\ \emph {et~al.}(2016)\citenamefont {Zhou},
  \citenamefont {Tolstikhin},\ and\ \citenamefont {Morishita}}]{Zhou_2016}%
  \BibitemOpen
  \bibfield  {author} {\bibinfo {author} {\bibfnamefont {Y.}~\bibnamefont
  {Zhou}}, \bibinfo {author} {\bibfnamefont {O.~I.}\ \bibnamefont
  {Tolstikhin}},\ and\ \bibinfo {author} {\bibfnamefont {T.}~\bibnamefont
  {Morishita}},\ }\bibfield  {title} {\bibinfo {title} {Near-forward
  rescattering photoelectron holography in strong-field ionization: Extraction
  of the phase of the scattering amplitude},\ }\href
  {https://doi.org/10.1103/PhysRevLett.116.173001} {\bibfield  {journal}
  {\bibinfo  {journal} {Phys. Rev. Lett.}\ }\textbf {\bibinfo {volume} {116}},\
  \bibinfo {pages} {173001} (\bibinfo {year} {2016})}\BibitemShut {NoStop}%
\bibitem [{\citenamefont {Liu}\ \emph {et~al.}(2016)\citenamefont {Liu},
  \citenamefont {Li}, \citenamefont {Wu}, \citenamefont {Gong}, \citenamefont
  {Staudte},\ and\ \citenamefont {Liu}}]{liu2016phase}%
  \BibitemOpen
  \bibfield  {author} {\bibinfo {author} {\bibfnamefont {M.-M.}\ \bibnamefont
  {Liu}}, \bibinfo {author} {\bibfnamefont {M.}~\bibnamefont {Li}}, \bibinfo
  {author} {\bibfnamefont {C.}~\bibnamefont {Wu}}, \bibinfo {author}
  {\bibfnamefont {Q.}~\bibnamefont {Gong}}, \bibinfo {author} {\bibfnamefont
  {A.}~\bibnamefont {Staudte}},\ and\ \bibinfo {author} {\bibfnamefont
  {Y.}~\bibnamefont {Liu}},\ }\bibfield  {title} {\bibinfo {title} {Phase
  structure of strong-field tunneling wave packets from molecules},\
  }\href@noop {} {\bibfield  {journal} {\bibinfo  {journal} {Phys. Rev. Lett.}\
  }\textbf {\bibinfo {volume} {116}},\ \bibinfo {pages} {163004} (\bibinfo
  {year} {2016})}\BibitemShut {NoStop}%
\end{thebibliography}%

\end{document}